\documentclass[12pt]{article}

\usepackage{graphics}
\usepackage{amsmath}
\usepackage{epsfig}
\usepackage{cite}

\usepackage[T2A]{fontenc} 
\usepackage{amssymb,bm,amsfonts,amsmath,textcomp,mathtext,indentfirst} 
\usepackage{misccorr,enumerate,cite,csquotes,tocloft,float,setspace,xcolor}
\usepackage{graphicx}
\usepackage[singlelinecheck=false]{caption}
\title{Bottomonia production and polarization in the NRQCD with $k_T$-factorization. I: $\Upsilon(3S)$ and $\chi_b(3P)$ mesons}
\author{N.A.~Abdulov$^{1}$, A.V.~Lipatov$^{2,\,3}$}

\begin{document}

\maketitle

\begin{center}

{\it $^1$Faculty of Physics, Lomonosov Moscow State University, 119991 Moscow, Russia}\\
{\it $^2$Skobeltsyn Institute of Nuclear Physics, Lomonosov Moscow State University, 119991 Moscow, Russia}\\
{\it $^3$Joint Institute for Nuclear Research, 141980 Dubna, Moscow Region, Russia}

\end{center} 

\vspace{0.2cm}

\begin{center}

{\bf Abstract }

\end{center}

The $\Upsilon(3S)$ production and polarization at high energies is 
studied in the framework of $k_T$-factorization approach.
Our consideration is based on the non-relativistic 
QCD formalism for bound states formation and off-shell production
amplitudes for hard partonic 
subprocesses. The transverse momentum dependent (TMD, or unintegrated) gluon
densities in a proton were derived from the 
Ciafaloni-Catani-Fiorani-Marchesini (CCFM)
evolution equation as well as from the Kimber-Martin-Ryskin (KMR) prescription. 
Treating the non-perturbative color octet transitions 
in terms of the multipole radiation theory and taking into account feed-down
contributions from radiative $\chi_b(3P)$ decays, we extract 
the corresponding non-perturbative matrix elements  
for $\Upsilon(3S)$ and $\chi_b(3P)$ mesons
from a combined fit to $\Upsilon(3S)$ transverse momenta distributions 
measured by the CMS and ATLAS Collaborations 
at the~LHC energies $\sqrt s = 7$ and $13$~TeV and 
central rapidities. 
Then we apply the extracted values to 
describe the CDF and LHCb data on $\Upsilon(3S)$
production and to investigate the polarization 
parameters $\lambda_\theta$, $\lambda_\phi$ and $\lambda_{\theta\phi}$, 
which determine the $\Upsilon(3S)$ spin density matrix. 
Our predictions
have a good agreement with the currently 
available data within the theoretical and experimental uncertainties. 

\newpage

\section{Introduction} \indent

Since it was first observed, the production of charmonia and 
bottomonia in hadronic collisions remains a 
subject of considerable theoretical and experimental studies.
A theoretical framework for the description 
of heavy quarkonia production and decays is provided by the
non-relativistic QCD (NRQCD) factorization\cite{1,2}. 
This formalism implies a separation of perturbatively 
calculated short-distance cross-sections for
the production of $Q\bar Q$ pair in an intermediate 
Fock state ${}^{2S+1}L_J^{(a)}$ with spin $S$, 
orbital angular momentum $L$, total angular momentum $J$ and 
color representation $a$ from long-distance 
non-perturbative matrix elements (NMEs), which describe the 
transition of that intermediate $Q\bar Q$ state 
into a physical quarkonium via soft gluon radiation. 
The NMEs are assumed to be universal 
(process- and energy-independent), not dependent on the quarkonium
momentum and obeying certain 
hierarchy in powers of the relative heavy quark velocity 
$v_Q \sim \log^{-1} m_Q/\Lambda_{\rm QCD}$ with $m_Q$ being 
the heavy quark mass. The color octet (CO) NMEs are not 
calculable within the theory and have to be only extracted from the data.

At present, the cross sections of prompt $S$- and $P$-wave 
charmonia production ($\psi^\prime$, $\chi_c$, $J/\psi$ and $\eta_c$)
in $pp$ collisions are known at the next-to-leading order
(NLO NRQCD)\cite{3,4,5,6,7,8,9,10,11,12,13,14,15}. 
The dominant tree-level
next-to-next-to-leading order (NNLO$^*$) corrections to 
the color-singlet (CS) production mechanism have been calculated\cite{16}.
With properly adjusted values of NMEs, one can achieve a good agreement between the NLO NRQCD predictions
and the experimental data on the $\psi^\prime$, $\chi_c$ 
and $J/\psi$ transverse momenta distributions\cite{3,4,5,6,7,8,9,10}.
However, the extracted NMEs strongly 
depend on the minimal charmonia transverse momentum $p_T$ used in the fits 
and are almost incompatible with 
each other when obtained from fitting different data sets. 
Moreover, none of the fits can reasonably 
describe the $\psi^\prime$ and $J/\psi$ polarization data 
(the so-called ``polarization puzzle''). The fits involving low-$p_T$ data 
result in the conclusion that the $\psi^\prime$ and $J/\psi$ production 
at large transverse momenta is
dominated by color-octet ${}^3S_1^{(8)}$ contributions with 
strong transverse polarization, that contradicts 
to the unpolarized production seen 
at the Tevatron and LHC.
To obtain an unpolarized $\psi^\prime$ and $J/\psi$ mesons, 
it is necessary  to assume that the production is dominated 
by the scalar ${}^1S_0^{(8)}$ intermediate state\cite{4}. 
However, such assumption immediately contradicts recent $\eta_c$ 
production data since the respective $\eta_c$ and $J/\psi$ NMEs are
related by the heavy quark spin symmetry (HQSS) principle\cite{1,2}. 
The HQSS requires that the $\eta_c$ and $J/\psi$ NMEs have to be 
determined from the simultaneous fit for the entire charmonia family, that
turned out to be impossible in the NLO NRQCD\footnote{The impact of the $\eta_c$ data on 
charmonia production and polarization was investigated\cite{12}.}
(see also discussions\cite{17,18,19}). 
The overall complicated situation is still far from understanding and 
has been even called ``challenging''\cite{13}.

A possible solution to the problem above has been proposed recently\cite{20} 
in the framework of a model 
that interprets the soft final state gluon radiation 
(which transforms an unbound $Q\bar Q$ pair into a 
physical quarkonium state) as a series of color-electric dipole transitions.
In this way the NMEs are represented in an explicit form inspired
by the classical multipole radiation theory, that has
dramatic consequences for the polarization of the final state mesons 
since the spin structure of the transition amplitudes is specified. 
The proposed approach leads 
to unpolarized or only weakly polarized charmonia either 
because of the cancellation 
between the ${}^3P_1^{(8)}$ and ${}^3P_2^{(8)}$ contributions
or as a result of two
successive color-electric $E1$ dipole transitions in the 
chain ${}^3S_1^{(8)} \rightarrow {}^3P_J^{(8)} \rightarrow {}^3S_1^{(1)}$ 
giving us the possibility to simultaneously solve 
the polarization puzzle for $J/\psi$ mesons and 
production puzzle for $\eta_c$ mesons\cite{21,22}.

An alternative laboratory for understanding the physics of the 
hadronization of heavy quark pairs is provided by the 
$\Upsilon(nS)$ and $\chi_b(mP)$ production, which has been measured
recently by the CMS\cite{23,24}, ATLAS\cite{25} and LHCb\cite{26,27} 
Collaborations at the LHC. Polarization of $\Upsilon(nS)$ 
mesons has been also investigated
by the CMS\cite{28} and LHCb\cite{29} Collaborations.
Due to heavier masses of bottomonia and smaller relative velocity $v_b$ of $b$
quarks in the bottomonium
rest frame ($v_b \simeq 0.08$ against $v_c \simeq 0.23$), these processes could be even a more suitable 
case to apply the NRQCD factorization because of a more faster convergence of the double NRQCD expansion in 
strong coupling $\alpha_s$ and $v_Q$. 
The complete NLO NRQCD predictions for $\Upsilon(nS)$ and $\chi_b(mP)$ 
production in $pp$ and $p\bar p$ collisions
were presented\cite{30,31,32,33,34}. As it was shown, one can 
reasonably explain the 
LHC data, both on $\Upsilon(nS)$ or $\chi_b(mP)$ 
yield and $\Upsilon(nS)$ polarization,
by taking into account 
latest measurements on the $\chi_b(mP)$ production. 
In particular, the polarization
puzzle for $\Upsilon(3S)$ meson can be solved by considering 
the $\chi_b(3P)$ feed-down contributions\cite{33,34}.
The latter have been observed recently by the LHCb Collaboration 
for the first time\cite{35} and found to be rather 
significant (up to 40\%). 

However, it is important to investigate 
the $S$- and $P$-wave bottomonia production and polarization 
within the same framework which has been already successfully applied 
for charmonia\cite{21,22}. 
Here we start with a short series of papers dedicated to the 
$\Upsilon(nS)$ and $\chi_b(mP)$ production in $pp$ and $p\bar p$ collisions at the high energies, that 
continues line of our previous studies.
In the present note we concentrate on the $\Upsilon(3S)$ production
with a consistent treatment for large $\chi_b(3P)$ feed-down contribution.
The $\Upsilon(1S), \Upsilon(2S), \chi_b(1P)$ and $\chi_b(2P)$ production 
requires a dedicated study which will be the subject of our forthcoming papers.
To describe the perturbative production of the $b\bar b$ pair in the 
hard scattering subprocess we apply the $k_T$-factorization  
approach\cite{36,37}. 
This approach is based on the Balitsky-Fadin-Kuraev-Lipatov 
(BFKL)\cite{38} or 
Ciafaloni-Catani-Fiorani-Marchesini (CCFM)\cite{39} evolution 
equations, which resum 
large logarithmic terms proportional to $\ln s \sim \ln 1/x$, important 
at high energies 
(or, equivalently, at low longitudinal momentum fraction $x$ of proton 
carried by gluon).
The $k_T$-factorization approach has certain technical advantages
in the ease of including higher-order radiative 
corrections (namely, leading part of NLO + NNLO + $\dots$ terms 
corresponding to real 
gluon emissions in initial state) in the form of 
transverse momentum dependent (TMD, or unintegrated) 
gluon density function in a proton\footnote{For different 
aspects of using the $k_T$-factorization
approach the reader may consult the review\cite{40}.}.
To describe the non-perturbative transition of an unbound $b\bar b$ 
pair into physical bottomonia
we employ the model\cite{20}.
We determine the NMEs for $\Upsilon(3S)$ and $\chi_b(3P)$ mesons 
from the $\Upsilon(3S)$ transverse momentum 
distributions measured by the CMS\cite{23,24} and ATLAS\cite{25} 
Collaborations in the central rapidity region at $\sqrt s = 7$ and $13$~TeV
(where the $k_T$-factorization approach is expected to 
be mostly relevant) and from the
relative production ratio $R^{\chi_b(3P)}_{\Upsilon(3S)}$
measured recently by the LHCb Collaboration at $\sqrt s = 7$ and 
$8$~TeV\cite{35}. 
Then, we examine the extracted 
NMEs on the Tevatron and LHC data taken by the 
CDF\cite{41} and LHCb\cite{26,27} Collaborations
and make predictions for polarization 
parameters $\lambda_\theta$, $\lambda_\phi$, $\lambda_{\theta\phi}$ 
(and frame-independent parameter $\tilde\lambda$), which determine 
the $\Upsilon(3S)$ spin density matrix and compare
them to the currently available data\cite{28,42}.

The outline of our paper is the following. In Section 2 
we briefly recall the basic steps of our calculations. 
In Section 3 we perform a numerical fit and extract the NMEs
from the LHC data. 
Then we check the compatibility of the extracted NMEs with the 
available data on $\Upsilon(3S)$ yeild and
polarization. Our conclusions are collected in Section 4.

\section{Theoretical framework} \indent

In the present note we follow the approach described in the 
earlier publications\cite{43,44,45}.
For the reader's convenience, we briefly recall here main 
points of the theoretical scheme.
Our consideration is based on the off-shell gluon-gluon fusion 
subprocesses that represent
the true leading order (LO) in QCD:
\begin{equation}
g^*(k_1) + g^*(k_2) \rightarrow \Upsilon[{}^3S_1^{(1)}](p) + g(k),
\end{equation}	
\begin{equation}
g^*(k_1) + g^*(k_2) \rightarrow \Upsilon[{}^1S_0^{(8)},{}^3S_1^{(8)},{}^3P_J^{(8)}](p).
\end{equation}	
\begin{equation}
g^*(k_1) + g^*(k_2) \rightarrow \chi_{bJ}(p)[{}^3P_J^{(1)},{}^3S_1^{(8)}] \rightarrow \Upsilon(p_1) + \gamma(p_2),
\end{equation}

\noindent where $J = 0, 1$ or $2$ and the four-momenta of all particles 
are given in the parentheses. The color states taken into account 
are directly indicated.
To obtain the production amplitudes for $b\bar b$ states with required 
quantum numbers from 
the ones for an unspecified $b\bar b$ state we use the appropriate 
projection operators. 
These operators for the spin-singlet and spin-triplet states 
read\cite{46}:
\begin{equation}
\Pi_0 = (\hat p_{\bar b} - m_b)\gamma_5(\hat p_{b} + m_b)/m^{3/2},
\end{equation}
\begin{equation}
\Pi_1 = (\hat p_{\bar b} - m_b)\hat\epsilon(S_z)(\hat p_{b} + m_b)/m^{3/2},
\end{equation}

\noindent where $m = 2m_b$, $p_b = p/2 + q$ and $p_{\bar b} = p/2 - q$ 
are the four-momenta of the quark and 
anti-quark and $q$ is the four-momentum of quarks in the bound state, 
which is associated with the orbital angular momentum $L$. 
States with various projections of the spin momentum onto the $z$ axis 
are represented by the 
polarization four-vector $\hat\epsilon(S_z)$.
Then, to calculate off-shell production amplitudes (1) --- (3), 
one has to integrate the product of the 
hard scattering amplitude $A(q)$ expanded in a series 
around $q = 0$ and meson bound state
wave function $\Psi^{(a)}(q)$ with respect to $q$:
\begin{equation}
A(q)\Psi^{(a)}(q) = A|_{q=0}\Psi^{(a)}(q) + q^\alpha(\partial A/\partial q^\alpha)|_{q=0}\Psi^{(a)}(q) + \dots
\end{equation}

\noindent A term-by-term integration of this series employs the 
identities\cite{46}:
\begin{equation}
\int{{{d^3q}\over{(2\pi)^3}}\Psi^{(a)}(q)} = {{1}\over{\sqrt{4\pi}}}\mathcal{R}^{(a)}(0),
\end{equation}	
\begin{equation}
\int{{{d^3q}\over{(2\pi)^3}}q^\alpha\Psi^{(a)}(q)} = -i\epsilon^\alpha(L_z){{\sqrt{3}}\over{\sqrt{4\pi}}}\mathcal{R'}^{(a)}(0),
\end{equation}

\noindent where $\mathcal{R}^{(a)}(x)$ is the radial wave function 
in the coordinate representation.
The first term in (6) contributes to $S$-waves only and 
vanishes for $P$-wave. 
In contrast, the second term contributes only 
to $P$-waves and vanishes for $S$-wave. 
States with various projections of the orbital 
angular momentum onto the $z$ axis are represented 
by the polarization four-vector $\epsilon_\mu(L_z)$. The 
corresponding NMEs are directly 
related to the wave functions $\mathcal{R}^{(a)}(x)$ 
and their derivatives\cite{1,2}:
\begin{equation}
\langle\mathcal{O}^{\cal Q}[{}^{2S+1}L_J^{(a)}]\rangle = 2N_c(2J+1)|\mathcal{R}^{(a)}(0)|^2/4\pi,
\end{equation}
\begin{equation}
\langle\mathcal{O}^{\cal Q}[{}^{2S+1}L_J^{(a)}]\rangle = 6N_c(2J+1)|\mathcal{R'}^{(a)}(0)|^2/4\pi
\end{equation}

\noindent for $S$- and $P$-wave quarkonium $\cal Q$ respectively, 
where $N_c = 3$. Additionally, the NMEs obey the multiplicity relations 
coming from HQSS at LO: 
\begin{equation}
\langle\mathcal{O}^{\cal Q}[{}^{3}P_J^{(a)}]\rangle = (2J+1)\langle\mathcal{O}^{\cal Q}[{}^{3}P_0^{(a)}]\rangle.
\end{equation}

\noindent
A similar relation holds for color-octet $^3S_1^{(8)}$ states if
$P$-wave quarkonia are considered.
The color-singlet wave functions and their derivatives can be obtained 
from the potential model calculation\cite{47,48} or extracted from 
the measured quarkonia decay widths.
Further evaluation of partonic amplitudes is straightforward and was
done in our previous papers\cite{43,44,45}. We only mention here that 
the summation over  
polarizations of the incoming off-shell gluons is performed 
according the BFKL prescription
${\sum {\epsilon^\mu\epsilon^{*\nu}}} = {\bm k}_T^\mu{\bm k}_T^\nu/{\bm k}_T^2$, 
where ${\bm k}_T$ is the gluon transverse momentum orthogonal 
to the beam axis\cite{36,37}. 
The spin density matrix of the $S$-wave quarkonia is expressed
in terms of the 
momenta $l_1$ and $l_2$ of the decay leptons and reads
\begin{equation}
{\sum {\epsilon^\mu\epsilon^{*\nu}}} = 3(l_1^\mu l_2^\nu + l_1^\nu l_2^\mu - {{m^2}\over{2}}g^{\mu\nu})/m^2.
\end{equation}

\noindent
This expression is equivalent to the standard expression 
$\sum{\epsilon^\mu\epsilon^{*\nu}} = - g^{\mu\nu} + p^\mu p^\nu/m^2$, but 
more suitable for 
determining the polarization observables. In all other respects 
the evaluation follows the standard QCD Feynman rules. 
The obtained results have been explicitly tested 
for gauge invariance by substituting 
the gluon momenta for corresponding polarization vectors. 
We have observed their gauge invariance even with off-shell initial 
gluons\footnote{Our results for perturbative production amplitudes 
squared and summed over polarization states agree with ones\cite{49}.}. 

As it was done for the prompt charmonia production\cite{21,22}, 
to describe the transition of an 
unbound octet $b\bar b$ quark pair to an observed singlet 
state we employ the mechanism proposed in\cite{20}. 
In this approach, a soft gluon with a small 
energy $E\sim\Lambda_{\rm QCD}$ is emitted after the hard 
interaction is over, bringing away the unwanted color and 
changing other quantum numbers of the 
produced CO system. In the conventional NRQCD calculations the
emitted final state gluons are regarded as carrying no energy-momentum, 
that is in obvious contradiction with confinement, which 
prohibits the emission 
of infinitely soft colored quanta.
In reality, the $b\bar b$ system must undergo a kind 
of final state interaction, 
where the energy-momentum exchange must be larger than 
the typical confinement scale.
Thus, having small energy of the emitted gluons gives us
the confidence that we do not enter the 
confinement or perturbative 
domains\footnote{The dependence of the numerical results 
on the emitted energy $E$ is discussed in Section 3.}. 
This is not the matter of only kinematical corrections 
since one cannot organize transition amplitudes 
with correct spin properties without some finite energy-momentum transfer.
In our calculations such soft gluon emission is described
by a classical multipole expansion, 
in which the electric dipole ($E1$) transition dominates\cite{50}. 
Only a single $E1$ transition is needed to transform a $P$-wave 
state into an $S$-wave state and the structure of the 
respective ${^3P_J^{(8)}}\to {^3S_1^{(1)}}+g$ amplitudes
is given by\cite{50}:
\begin{equation}
A({}^3P_0^{(8)} \rightarrow \Upsilon + g) \sim k_\mu^{(g)}p^{\rm (CO)\mu}\epsilon_\nu^{(\Upsilon)}\epsilon^{(g)\nu},
\end{equation}
\begin{equation}
A({}^3P_1^{(8)} \rightarrow \Upsilon + g) \sim e^{\mu\nu\alpha\beta}k_\mu^{(g)}\epsilon_\nu^{\rm (CO)}\epsilon_\alpha^{(\Upsilon)}\epsilon_\beta^{(g)},
\end{equation}
\begin{equation}
A({}^3P_2^{(8)} \rightarrow \Upsilon + g) \sim p_\mu^{\rm (CO)}\epsilon_{\alpha\beta}^{\rm (CO)}\epsilon_\alpha^{(\Upsilon)} \left[ k_\mu^{(g)}\epsilon_\beta^{(g)} - k_\beta^{(g)}\epsilon_\mu^{(g)} \right],
\end{equation}

\noindent where $p^{\rm (CO)}_\mu$, $k^{(g)}_\mu$, 
$\epsilon^{(\Upsilon)}_\mu$, $\epsilon^{(g)}_\mu$, 
$\epsilon^{\rm (CO)}_\mu$ and $\epsilon^{\rm (CO)}_\mu$ are the 
momenta and polarization vectors of corresponding particles 
and $e^{\mu\nu\alpha\beta}$ is the fully 
antisymmetric Levi-Civita tensor. 
The transformation of color-octet $S$-wave state into the
color-singlet $S$-wave state is treated as 
two successive $E1$ transitions ${}^3S_1^{(8)} \rightarrow {}^3P_J^{(8)} + g$, 
${}^3P_J^{(8)} \rightarrow {}^3S_1^{(1)} + g$ proceeding via either of 
three intermediate ${}^3P_J^{(8)}$ states with $J = 0,1,2$. 
For each of these transitions 
we apply the same expressions (13) --- (15). Of course, all the
expressions above are the same 
for gluons and photons (up to an overall color factor) and 
therefore can be used to 
calculate the polarization variables in radiative decays
in feed-down 
process $\chi_b(3P) \rightarrow \Upsilon(3S) + \gamma$. 
Thus, the polarization of the outgoing $\Upsilon(3S)$ 
meson can then be calculated without any ambiguity.

The approach\cite{20} contrasts to conventional NRQCD 
calculations which show that heavy
quarkonia produced from high-$p_T$ gluons as $^3S_1^{(8)}$ states
carry strong transverse polarization. 
With our completely different view on the heavy quarkonia 
depolarization mechanism, we 
finally arrive at a completely different set of the 
fitted NMEs (see Section~3).
The squares of the matrix elements, as being too lengthy, 
are not presented here
but implemented into the newly developed parton-level Monte-Carlo event 
generator \textsc{pegasus}\cite{51}.

The cross sections of $\Upsilon(3S)$ and $\chi_{b}(3P)$ production 
in the $k_T$-factorization 
approach are calculated as a convolution of the off-shell 
partonic cross sections and 
TMD gluon densities $f_g(x,\bm{k}^2_T,\mu^2)$ in a proton. 
The cross section 
for $2 \rightarrow 2$ and $2 \rightarrow 1$ subprocesses (1) --- (3) 
can be written as:
\begin{multline}
\sigma = \int{{1}\over{8\pi(x_1x_2s) F}}f_g(x_1,\bm{k}^2_{1T},\mu^2)f_g(x_2,\bm{k}^2_{2T},\mu^2)\times\\
\times \overline{|A(g^*+g^*\rightarrow Q\bar Q + g)|^2}d\bm{p}^2_{T}d\bm{k}^2_{1T}d\bm{k}^2_{2T}dydy_g{{d\phi_1}\over{2\pi}}{{d\phi_2}\over{2\pi}},
\end{multline}	
\begin{equation}
\sigma=\int{{{2\pi}\over{x_1x_2sF}}f_g(x_1,\bm{k}^2_{1T},\mu^2)f_g(x_2,\bm{k}^2_{2T},\mu^2)\overline{|A(g^*+g^*\rightarrow Q\bar Q)|^2}}d\bm{k}^2_{1T}d\bm{k}^2_{2T}dy{{d\phi_1}\over{2\pi}}{{d\phi_2}\over{2\pi}},
\label{sigma}
\end{equation} 

\noindent where $\phi_1$ and $\phi_2$ are the azimuthal angles of the 
initial off-shell gluons having 
the fractions of the momentum $x_1$ and $x_2$ and non-zero transverse 
momenta $\bm{k}^2_{1T}$ and $\bm{k}^2_{2T}$, $\bm{p}^2_{T}$ and $y$ are the transverse 
momentum and rapidity of produced mesons, $y_g$ is the rapidity of the
outgoing gluon and $\sqrt{s}$ is the $pp$ center-of-mass energy.
According to the general definition\cite{52},
the off-shell gluon flux factor $F$ is defined as 
$F = 2\lambda^{1/2}(\hat s, k_1^2, k_2^2)$, 
where $\hat s = (k_1 + k_2)^2$ and $\lambda(x,y,z)$ is the known 
kinematic function.
Note that for $2 \to 2$ subprocesses one can use the approximation
$\lambda^{1/2}(\hat s, k_1^2, k_2^2) \simeq \hat s \simeq x_1 x_2 s$.
However, it is not suitable for the $2 \to 1$ kinematics 
because the difference between $\hat s \simeq m^2_\Upsilon$
and $x_1 x_2 s = m_\Upsilon^2 + p_T^2$ can make pronounced 
effect on the $p_T$ spectrum.
This effect is specially discussed in Section~3.

In the present paper we have tested a few sets of the TMD gluon 
densities in a proton, 
namely, A0\cite{53}, JH'2013~set~1\cite{54} and KMR\cite{55} 
ones\footnote{A comprehensive collection of the TMD gluon distributions 
can be found in the \textsc{tmdlib} package\cite{56}, which is 
a C++ library providing a 
framework and an interface to the different parametrizations.}. 
First two of them were obtained from the numerical solutions of 
the CCFM gluon evolution equation.
The CCFM equation provides a suitable tool since it converges to the
BFKL equation in the
region of small $x$ and to the DGLAP equation at large $x$ (see\cite{39} for 
more details). 
The typical values of the variable $x$ probed in the considered processes 
are of order $x \sim (m_\Upsilon^2 + p_T^2)^{1/2}/\sqrt s$ at central 
rapidities,
that corresponds to $x \sim 10^{-3} \, ... \, 10^{-2}$ in the kinematical
conditions of the CMS and ATLAS experiments\cite{23,24,25}. Thus, 
the CCFM evolution 
can be used
in the whole $p_T$ range.
The input parameters of these gluon distributions were 
determined from the best description of the precision DIS data on the 
proton structure functions $F_2(x, Q^2)$. 
Additionally, we have used a set obtained with Kimber-Martin-Ryskin (KMR) 
prescription\cite{55}, which provides a method to evaluate the
TMD parton densities from the conventional (collinear) ones. 
For the input, we have used recent LO 
NNPDF3.1 set\cite{57}.
The A0, JH'2013 and KMR gluon densities are shown in Fig.~1
as a function of ${\mathbf k}_T^2$ for different values of $x$
and $\mu^2$. One can observe a difference in the absolute 
normalization and shape between all these TMD gluon distributions. 
Below we discuss the corresponding phenomenological consequences.

The renormalization $\mu_R$ and factorization $\mu_F$ scales were set to 
$\mu_R^2 = m_\Upsilon^2 + p_T^2$ and 
$\mu_F^2 = \hat s + {\mathbf Q}_T^2$ for CCFM-evolved gluon densities, 
where ${\mathbf Q}_T$ is the transverse momentum of the initial off-shell 
gluon pair. The choice of $\mu_R$ is a standard for 
bottomonia production, while the special choice of $\mu_F$ is 
connected with the CCFM evolution (see\cite{53,54}). 
In the KMR calculations, we used standard choice 
$\mu_R^2 = \mu_F^2 = m_\Upsilon^2 + p_T^2$.
       
The parton level calculations were performed using the 
Monte-Carlo event generator \textsc{pegasus}\cite{51}. 	

\section{Numerical results} \indent

As it was mentioned above, in the present paper we concentrate on
the inclusive $\Upsilon(3S)$ and $\chi_b(3P)$ production, leaving other 
bottomonia states for forthcoming studies.
Below we set the masses  
$m_{\Upsilon(3S)} = 10.3552$~GeV, $m_{\chi_{b1}(3P)} = 10.512$~GeV and 
$m_{\chi_{b2}(3P)} = 10.522$~GeV~\cite{58} and adopt the 
usual non-relativistic approximation $m_b = m_{\cal Q}/2$
for the beauty quark mass, where $m_{\cal Q}$ is the mass of 
bottomonium $\cal Q$.
We set the branching ratios $B(\Upsilon(3S) \rightarrow \mu^+\mu^-) = 0.0218$\cite{58}, 
$B(\chi_{b1}(3P)\rightarrow \Upsilon(3S) + \gamma) = 0.1044$ and 
$B(\chi_{b2}(3P) \rightarrow \Upsilon(3S) + \gamma) = 0.0611$\cite{34}. 
Note that there are no experimental data for branching 
ratios of $\chi_b(3P)$,
so the values above are the results of an assumption\cite{34}
that the total decay widths of $\chi_b(mP)$ are approximately 
independent on $m$.
Following experimental analysis\cite{35}, we 
neglected the $\chi_{b0}(3P)$ contribution as it is almost zero.
We use the one-loop formula for the coupling $\alpha_s$ 
with $n_f = 4(5)$ quark flavours at $\Lambda_{\rm QCD} = 250(167)$~MeV 
for A0 (KMR) gluon density
and two-loop expression for $\alpha_s$ with $n_f = 4$ 
and $\Lambda_{\rm QCD} = 200$~MeV for JH'2013 set 1 gluon.  
As a commonly adopted choice, we set 
CS NMEs $\langle\mathcal{O}(\Upsilon[{}^{3}S_1^{(1)}])\rangle  = 3.54$ GeV$^3$ 
and $\langle\mathcal{O}(\chi[{}^{3}P_0^{(1)}])\rangle  = 2.83$ GeV$^5$.
These values were obtained in the potential model calculations\cite{47}.

\subsection{Fit of color octet NMEs} \indent

We have performed a global fit to the $\Upsilon(3S)$ production data at 
the 
LHC and determined the corresponding NMEs for 
both $\Upsilon(3S)$ and $\chi_b(3P)$ mesons.
We have included in the fitting procedure the $\Upsilon(3S)$ transverse 
momentum distributions measured by
the CMS\cite{23,24} and ATLAS\cite{25} Collaborations 
at $\sqrt s = 7$ and $13$~TeV and central rapidities, where 
our $k_T$-factorization calculations are most relevant due to 
essentially low-$x$ region probed.
To determine NMEs for $\chi_b(3P)$ mesons, we also included into the fit 
the 
recent LHCb data\cite{35} on the radiative 
$\chi_b(3P) \to \Upsilon(3S) + \gamma$ decays 
taken at $\sqrt s = 7$ and $8$~TeV. 
We have excluded from our fit low $p_T$ region and consider 
only the data at
$p_T > p_T^{\rm cut} = 10$~GeV, where the NRQCD 
formalism is believed to be mostly reliable. 
As it was already mentioned above, 
the double NRQCD expansion in $\alpha_s$ and $v_Q$ is not good 
at low $p_T$, where a more accurate treatment of large
logarithms $\sim \ln m_{\Upsilon}^2/p_T^2$
and other nonperturbative effects becomes 
necessary\footnote{By this reason, we have also excluded from the fit  
the earlier CDF data\cite{41}, which mostly refer to the low 
$p_T$ region.}.

Before we proceed with the numerical fit, we would like to point 
out a few points.
First of them is connected with the importance of proper 
definition of the off-shell flux factor  
for $2 \to 1$ subprocesses (2) and (3).
The definition of the flux, which is the velocity of the 
off-shell interacting partons, is not clear and can be disputable.
As it was mentioned above, 
we use the ``$\lambda^{1/2}$'' prescription 
$F = 2\lambda^{1/2}(\hat s, k_1^2, k_2^2)$ in 
factorization formula (\ref{sigma}). Our choice 
is based on the toy simulation\cite{59} of $\chi_c$ 
meson production in $e^+e^-$ collisions.
It was argued\cite{59} that such definition leads to a good agreement
of calculations based on Equivalent Photon Approximation and 
exact ${\cal O}(\alpha^4)$ results.
Contrary, the calculations performed with using 
conventional (collinear) $2 \to 1$ flux treatment 
$\lambda^{1/2}(\hat s, k_1^2, k_2^2) \simeq x_1 x_2 s$ did not 
reproduce the latter and 
therefore, in our opinion, seems to be rather 
doubtful\footnote{Such calculations were done\cite{60}.}. 
        
Our calculation shows that 
the ``$\lambda^{1/2}$'' prescription results in different $p_T$ 
shapes of color-octet $^1S_0^{(8)}$ and $^3P_J^{(8)}$ contributions
to the $\Upsilon(3S)$ production. Let us consider 
the ratio $R$ defined as
\begin{equation}
R = { m_{\Upsilon(3S)}^2 \sum\limits_{J = 0}^{2} (2J+1) \, d\sigma[\Upsilon(3S), {}^3P_J^{(8)}]/dp_T \over d\sigma [\Upsilon(3S), {}^1S_0^{(8)}]/dp_T}
\label{eqrf}
\end{equation}

\noindent
as a function of $\Upsilon(3S)$ meson transverse momentum.
While the calculations with collinear treatment of the flux factor $F$ 
show a flat behavior of this ratio
in a wide $p_T$ region 
$10 < p_T < 100$~GeV (see Fig.~2, left panel)
the calculations performed with using the ``$\lambda^{1/2}$'' 
prescription demonstrate
the strong rise of the ratio $R$ with increasing $p_T$
giving us a possibility to separately extract the values of
$\langle\mathcal{O}^{\Upsilon(3S)}[{}^{1}S_0^{(8)}]\rangle$ and 
$\langle\mathcal{O}^{\Upsilon(3S)}[{}^{3}P_0^{(8)}]\rangle$ 
from the experimental data. 
The latter turns out to be impossible when one  
inconsistently uses the collinear treatment of 
flux factor in the $k_T$-factorization calculations.

Our next point is connected with the correct treatment of feed-down
contributions
from the radiative decays of $\chi_b(3P)$ mesons, observed 
recently by the LHCb Collaboration\cite{35}.
We found that the $p_T$ shape of the direct $\Upsilon[^3S_1^{(8)}]$ and 
feed-down $\chi_b[^3S_1^{(8)}]$ contributions is almost 
the same in all kinematical regions
probed by the LHC and Tevatron experiments. Thus, the ratio
\begin{equation}
r = { \sum\limits_{J = 0}^{2} (2J+1) \, B(\chi_{bJ}(3P) \to \Upsilon(3S) + \gamma) d\sigma[\chi_{bJ}(3P), {}^3S_1^{(8)}]/dp_T \over d\sigma [\Upsilon(3S), {}^3S_1^{(8)}]/dp_T }
\label{eqr}
\end{equation}

\noindent 
can be well approximated by a constant for a wide $\Upsilon(3S)$ 
transverse 
momentum $p_T$ and rapidity $y$ ranges at different energies, 
as it is demonstrated in Fig.~2 (right panel). 
We estimate the mean-square average $r = 0.654 \pm 0.005$, which is 
practically independent on the TMD gluon density in a proton.
Since up to now there are no experimental data on the $\chi_b(3P)$ 
transverse momentum distributions,
we cannot separately determine the values of 
$\langle\mathcal{O}^{\Upsilon(3S)}[{}^{3}S_1^{(8)}]\rangle$ and 
$\langle\mathcal{O}^{\chi_{b0}(3P)}[{}^{3}S_1^{(8)}]\rangle$ from the
available $\Upsilon(3S)$ data\cite{23,24,25}.
Instead, we introduce the linear combination
\begin{equation}
M_r = \langle\mathcal{O}^{\Upsilon(3S)}[{}^{3}S_1^{(8)}]\rangle + r \langle\mathcal{O}^{\chi_{b0}(3P)}[{}^{3}S_1^{(8)}]\rangle,
\end{equation}
\noindent
which can be extracted from the measured $\Upsilon(3S)$ transverse
momentum distributions.
Then we use recent LHCb data\cite{35} on the fraction
of $\Upsilon(3S)$ mesons 
originating from the $\chi_b(3P)$ radiative decays measured
at $\sqrt s = 7$ and $8$~TeV. 
To be precise, the LHCb Collaboration reported the ratio
\begin{equation}
R^{\chi_b(3P)}_{\Upsilon(3S)} = \sum\limits_{J = 1}^{2} {\sigma(pp\rightarrow \chi_{bJ}(3P)+X) \over \sigma(pp \rightarrow \Upsilon(3S) + X)} \times B(\chi_{bJ} \to \Upsilon(3S) + \gamma),
\label{eqrUp}
\end{equation}

\noindent
where the possible contributions from $\chi_{b0}(3P)$ decays are 
neglected because of the small 
branching fraction.
From the known $M_r$ and $R^{\chi_b(3P)}_{\Upsilon(3S)}$ values
one can separately determine the 
$\langle\mathcal{O}^{\Upsilon(3S)}[{}^{3}S_1^{(8)}]\rangle$ and 
$\langle\mathcal{O}^{\chi_{b0}(3P)}[{}^{3}S_1^{(8)}]\rangle$, thus
reconstructing full map of color octet NMEs for both $\Upsilon(3S)$ 
and $\chi_b(3P)$ mesons.

Using the strategy described above, we performed a numerical 
fit of $\Upsilon(3S)$ and $\chi_b(3P)$ NMEs.
Nowhere we impose any kinematic restrictions but the experimental 
acceptance.
The fitting procedure was separately done in each of the rapidity 
subdivisions (using the fitting algorithm as implemented 
in the commonly used \textsc{gnuplot} package\cite{61}) 
under the requirement that all the NMEs are strictly positive. 
Then, the mean-square average of
the fitted values was taken. The corresponding uncertainties are 
estimated in the conventional way
using Student's t-distribution at the confidence level $P = 80$\%.
The results of our fits are collected in Table~1. For comparison, 
we also presented 
there the NMEs obtained in the conventional NLO NRQCD by other 
authors\cite{33}.
We have found that extracted values 
of $\langle\mathcal{O}^{\Upsilon(3S)}[{}^{1}S_0^{(8)}]\rangle$ 
are compatible with zero for all the  
TMD gluon densities.
However, other color octet NMEs strongly depend on the 
latter, although JH'2013 set 1 and KMR gluons
result in the more or less close values.
The dependence of the fitted NMEs values on the TMD gluon densities 
reflects their different $x$ and ${\mathbf k}_T^2$ 
behavior, that is the consequence
of different approaches to evaluate them.
The corresponding $\chi^2/d.o.f.$ 
are listed in Table~2, where we additionally
show their dependence on the $p_T^{\rm cut}$.
As one can see, the $\chi^2/d.o.f.$ decreases when 
$p_T^{\rm cut}$ grows up and the
best fit of the data is achieved with A0 gluon.
We note that the returned relatively large (but still reasonable)
$\chi^2/d.o.f.$ values are connected with the recent 
precision CMS data\cite{24} included into the fit.
So, as an exercise, we have excluded these data and repeated the fit 
procedure using the ATLAS data\cite{25} only. In this 
way, the $\chi^2/d.o.f. \sim 1$ was obtained for 
all the considered TMD gluon densities.
    
All the data used in the fits are compared with our predictions 
in Figs.~3 --- 5. Note that the data at $p_T > p_T^{\rm cut} = 10$~GeV
are only shown.
The shaded areas represent the theoretical uncertainties of our 
calculations, 
which include the scale uncertainties, uncertainties coming from the NME 
fitting procedure and uncertainties connected with the 
choice of the intermediate color-octet mass, 
added in quadrature. 
To estimate the scale uncertainties
the standard variations in the scale 
$\mu_R\to 2\mu_R$ or $\mu_R\to\mu_R/2$ were introduced 
through replacing the gluon 
densities A0 and JH'2013 set 1 with A0$+$ and JH'2013 set 1$+$, or 
with A0$-$ and JH'2013 set 1$-$.
This was done to preserve the intrinsic correspondence between 
the TMD set and the scale used in the evolution equation (see\cite{53,54}).
To estimate the uncertainties connected with the 
intermediate color-octet mass we have varied 
amount of energy $E$ emitted in the course of transition of an
unbound color octet $b\bar b$ pair into the observed 
bottomonium by a factor of $2$
around its default value $E = \Lambda_{\rm QCD}$.
We find that the main effect here is only in changing the overall 
normalization with almost no changes in the shape of 
the $p_T$ spectrum (see also\cite{62}).
These uncertainties are about of $20$\% and therefore
comparable with the scale uncertainties. 
One can see that we have achieved a reasonably good 
description of the CMS\cite{23,24} and ATLAS\cite{25} 
data in the whole $p_T$ range within the experimental and 
theoretical uncertainties for the $\Upsilon(3S)$ transverse 
momentum distributions. The ratio $R^{\chi_b(3P)}_{\Upsilon(3S)}$
measured by the LHCb Collaboration\cite{35} is well described also.
At large $p_T$, the JH'2013 set 1 and KMR gluons tend to 
overestimate the latest CMS data\cite{24} taken at $\sqrt s = 13$~TeV,
but agree well with other measurements. 
This result shows a dependence of our predictions
on the TMD gluon densities in certain kinematical regions. 	
    
With obtained NMEs for $\Upsilon(3S)$ and $\chi_b(3P)$ mesons, we 
achieved reasonably good 
description (of course, at $p_T > p_T^{\rm cut}$) of 
the earlier CDF data\cite{41} taken at the $\sqrt s = 1.8$~TeV and
recent data\cite{26,27} taken by the LHCb Collaboration 
at $\sqrt s = 7$, $8$ 
and $13$~TeV and forward rapidities, see Fig.~6.
We find that the KMR gluon distribution is able to 
describe well the CDF data
even at low $p_T$ region, $p_T < 10$~GeV.
Some discrepancy between the LHCb 
data and our predictions observed in very forward 
region $4 < y < 4.5$
at $\sqrt s = 7$~TeV
can be easily understood since 
here one can probe the essentially large-$x$ region,
there the $k_T$-factorization becomes less applicable.

The consequence of our fit for $\Upsilon(3S)$ polarization 
is discussed in the next Section.
        
\subsection{$\Upsilon(3S)$ polarization} \indent

As it is well known, the polarization of any vector meson can be described 
with three parameters $\lambda_\theta$, $\lambda_\phi$ and 
$\lambda_{\theta\phi}$, which determine the spin density matrix of a 
meson decaying into a lepton pair and can be measured experimentally. 
The double differential angular distribution of the decay leptons 
can be written as\cite{63}:	
\begin{equation}
  {{d\sigma}\over{d\cos\theta^*d\phi^*}} \sim {{1}\over{3+\lambda_\theta}}(1 + \lambda_\theta\cos^2\theta^* + \lambda_\phi\sin^2\theta^*\cos2\phi^* + \lambda_{\theta\phi}\sin2\theta^*\cos\phi^*), 
  \label{eqlam}
\end{equation}

\noindent where $\theta^*$ and $\phi^*$ are the polar and azimuthal 
angles of the decay lepton measured in the meson rest frame.
The case of 
$(\lambda_\theta$, $\lambda_\phi$, $\lambda_{\theta \phi}) = (0,0,0)$
corresponds to unpolarized state, while 
$(\lambda_\theta$, $\lambda_\phi$, $\lambda_{\theta \phi}) = (1,0,0)$
and $(\lambda_\theta$, $\lambda_\phi$, $\lambda_{\theta \phi}) = (-1,0,0)$
refer to fully transverse and fully longitudinal polarizations. 

The CMS Collaboration has measured all of these parameters 
as functions of $\Upsilon(3S)$ transverse momentum 
in three complementary frames: the Collins-Soper, helicity and 
perpendicular helicity ones at $\sqrt{s} = 7$ TeV\cite{28}. 
The CDF Collaboration also measured these parameters  
in the helicity frame at $\sqrt{s} = 1.96$~TeV\cite{42}. 
In the Collins-Soper frame the polarization axis $z$ bisects
the two beam directions whereas the polarization axis in the 
helicity frame coincides with the $\Upsilon(3S)$ direction 
in the laboratory frame. 
In the perpendicular helicity frame the $z$ axis is orthogonal to that in the 
Collins-Soper frame and lies in the plane spanned by
the two beam ($P_1$ and $P_2$) momenta.
In all cases, the $y$ axis is taken to be in the direction of the 
vector product of the 
two beam directions in the $\Upsilon(3S)$
rest frame, $\vec P_1 \times \vec P_2$ and
$\vec P_2 \times \vec P_1$ for positive and negative rapidities, respectively. 
Additionally, the frame-independent parameter
$\tilde \lambda = (\lambda_\theta + 3\lambda_\phi)/(1 - \lambda_\phi)$ 
has been studied\cite{28,42}.
Below we estimate the polarization parameters 
$\lambda_\theta$, $\lambda_\phi$, $\lambda_{\theta \phi}$ and 
$\tilde \lambda$ for the CMS and CDF 
conditions\footnote{The LHCb Collaboration has also measured 
$\Upsilon(3S)$ polarization\cite{29}. However, these data were 
obtained at rather low transverse momenta and, therefore, we will not 
analyze them here.}. 
As it was done earlier\cite{22,43,44,45}, our calculation generally follows the
experimental procedure. We collect the simulated events in the 
kinematical region defined by the CMS and CDF experiments, 
generate the decay lepton angular distributions according
to the production and decay matrix elements and then 
apply a three-parametric fit based on (22).
Of course, we took into account the polarization of $\Upsilon(3S)$ 
mesons originated from radiative $\chi_b(3P)$ decays, 
that is in full agreement with the experimental setup.

Our results are presented in Figs.~7 --- 10. 
These calculations were done using the A0 
gluon density which provides the best description of the measured 
$\Upsilon(3S)$ transverse momenta distributions.
The obtained predictions for the $\Upsilon(3S)$ polarization 
parameters have a reasonable agreement with the CMS and CDF data. 
In all the kinematical regions we find only weak or zero polarization, 
which coincides with the measurements within the uncertainties. 
These predictions are practically independent of the 
$\Upsilon(3S)$ rapidity.
The absence of strong polarization is not connected with parameter 
tuning, but seems to be a natural and rather general 
feature of the scenario\cite{20}.
Thus, one can conclude that 
treating the soft gluon emissions within the NRQCD as a series 
of color-electric dipole transitions 
does not contradicts the available Tevatron and LHC data
on the $\Upsilon(3S)$ production.
The same conclusion was done for charmonia family\cite{21,22}.

Finally, we would like to note that the qualitative 
predictions for the $\lambda_\theta$, $\lambda_\phi$, 
$\lambda_{\theta\phi}$ and $\tilde \lambda$
are stable with respect to variations in the model parameters. 
In fact, there is practically no dependence on the strong coupling constant 
and TMD gluon densities, i.e. two of important sources of 
theoretical uncertainties cancel out.
So, the proposed way, in our opinion,
can provide an easy and natural solution to the 
quarkonia production and polarization puzzle.

\section{Conclusion} \indent

We have considered the $\Upsilon(3S)$ production at the Tevatron and LHC
in the framework of $k_T$-factorization 
approach. Our consideration was based on the off-shell 
production amplitudes for hard partonic subprocesses 
(including both color-singlet and color-octet contributions),  
NRQCD formalism for the formation of bound states
and TMD gluon densities in a proton (derived 
from the CCFM evolution equation and KMR scheme as well).
Treating the nonperturbative color octet transitions in terms of 
multipole radiation theory and
taking into account feed-down contributions from the 
radiative $\chi_b(3P)$ decays,
we extracted $\Upsilon(3S)$ and $\chi_b(3P)$ NMEs 
in a fit to $\Upsilon(3S)$ transverse momentum distributions 
measured by the CMS and ATLAS Collaborations at  
$\sqrt s = 7$ and $13$~TeV.
We have inspected the extracted NMEs with the available 
Tevatron and LHC data taken in different kinematical regions and
demostrated that these NMEs do not contradict the data.
We found that the best description is achieved with the CCFM-evolved A0 gluon
density, although the KMR one is able to describe the 
data even at low transverse momenta.
Then we
estimated polarization parameters $\lambda_\theta$, $\lambda_\phi$,  
$\lambda_{\theta \phi}$ and frame-independent parameter $\tilde \lambda$
which determine the $\Upsilon(3S)$ spin density matrix.
We show that treating the soft gluon emission as a 
series of explicit color-electric 
dipole transitions within the NRQCD leads to unpolarized 
$\Upsilon(3S)$ production at moderate and large transverse 
momenta, that is in agreement with the Tevatron and LHC data.

\section*{Acknowledgements} \indent

The authors thank S.P.~Baranov, M.A.~Malyshev and H.~Jung for their 
interest, useful discussions and important remarks.
N.A.A. is supported by the Foundation for the Advancement of 
Theoretical Physics and Mathematics ``Basis'' (grant No.18-1-5-33-1)
and by the RFBR grant 19-32-90096.
A.V.L. is grateful the DESY Directorate for the support
in the framework of Cooperation Agreement between MSU and DESY 
on phenomenology of the LHC processes and TMD parton densities.

	
	\newpage
	
	\begin{table}[H] \footnotesize
	\centering
	\begin{tabular}{lcccc}
	\hline
	\hline
	\\
	 &  A0 & JH'2013 set 1 &  KMR & NLO NRQCD\cite{33}\\
	\\ 
	\hline
	\\
	$\langle\mathcal{O}^{\Upsilon(3S)}[{}^{3}S_1^{(1)}]\rangle$/GeV$^{3}$ & $3.54$ & $3.54$ & $3.54$ & $3.54$ \\
    \\
	$\langle\mathcal{O}^{\Upsilon(3S)}[{}^{1}S_0^{(8)}]\rangle$/GeV$^3$ & $0.0$ & $0.0$ & $0.0$ & $-0.0107 \pm 0.0107$ \\	
	\\
	$\langle\mathcal{O}^{\Upsilon(3S)}[{}^{3}S_1^{(8)}]\rangle$/GeV$^3$ & $0.018 \pm 0.001$ & $0.007 \pm 0.002$ & $0.006 \pm 0.001$ & $0.0271 \pm 0.0013$ \\	
	\\
	$\langle\mathcal{O}^{\Upsilon(3S)}[{}^{3}P_0^{(8)}]\rangle$/GeV$^{5}$ & $0.0$ & $0.09 \pm 0.03$ & $0.073 \pm 0.006$ & $0.0039 \pm 0.0023$ \\	
	\\
	$\langle\mathcal{O}^{\chi_{b0}(3P)}[{}^{3}P_0^{(1)}]\rangle$/GeV$^{5}$ & $2.83$ & $2.83$ & $2.83$ & $2.83$ \\
    \\
	$\langle\mathcal{O}^{\chi_{b0}(3P)}[{}^{3}S_1^{(8)}]\rangle$/GeV$^{3}$ & $0.016 \pm 0.003$ & $0.009 \pm 0.001$ & $0.005 \pm 0.001$ & --- \\
    \\
	\hline
	\hline
	\end{tabular}
	\caption{The NMEs for $\Upsilon(3S)$ and $\chi_b(3P)$ mesons as determined from our 
	fit at $p_T^{\rm cut} = 10$~GeV. The NMEs obtained in the NLO NRQCD\cite{33} are shown for comparison.}
	\label{tab1}
	\end{table}
	
 	\begin{table}[H] \footnotesize
	\centering
	\begin{tabular}{lcccc}
	\hline
	\hline
	\\
	 & $p_T^{\rm cut} = 10$~GeV & $p_T^{\rm cut} = 12$~GeV & $p_T^{\rm cut} = 15$~GeV & $p_T^{\rm cut} = 17$~GeV\\
	\\ 
	\hline
	\\
	A0 & $2.35$ & $1.99$ & $1.79$ & $1.72$ \\	
	\\
	JH'2013 set 1 & $4.22$ & $3.59$ & $3.28$ & $3.21$ \\	
	\\
	KMR & $2.59$ & $2.43$ & $2.38$ & $2.37$ \\	
	\\
	\hline
	\hline
	\end{tabular}
	\caption{The dependence of the $\chi^2/d.o.f.$ achieved in the 
	fit procedure on the choice of $p_T^{\rm cut}$.}
	\label{tab2}
	\end{table}
	
\newpage	

\begin{figure}
\begin{center}
\includegraphics[width=7.0cm]{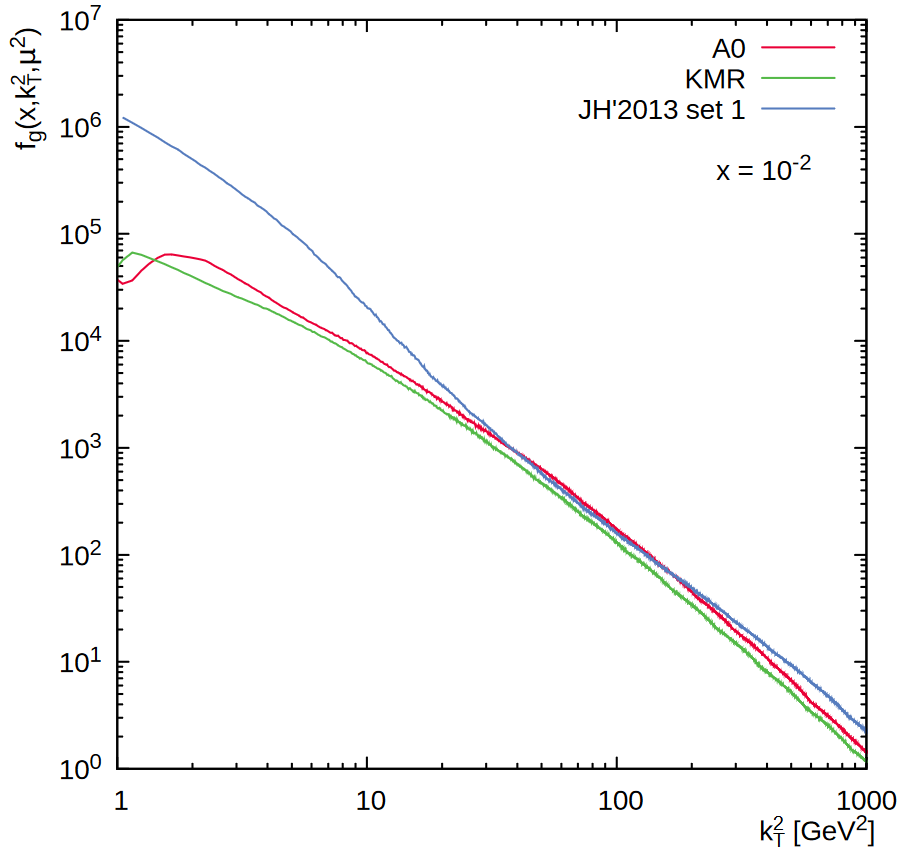}
\includegraphics[width=7.0cm]{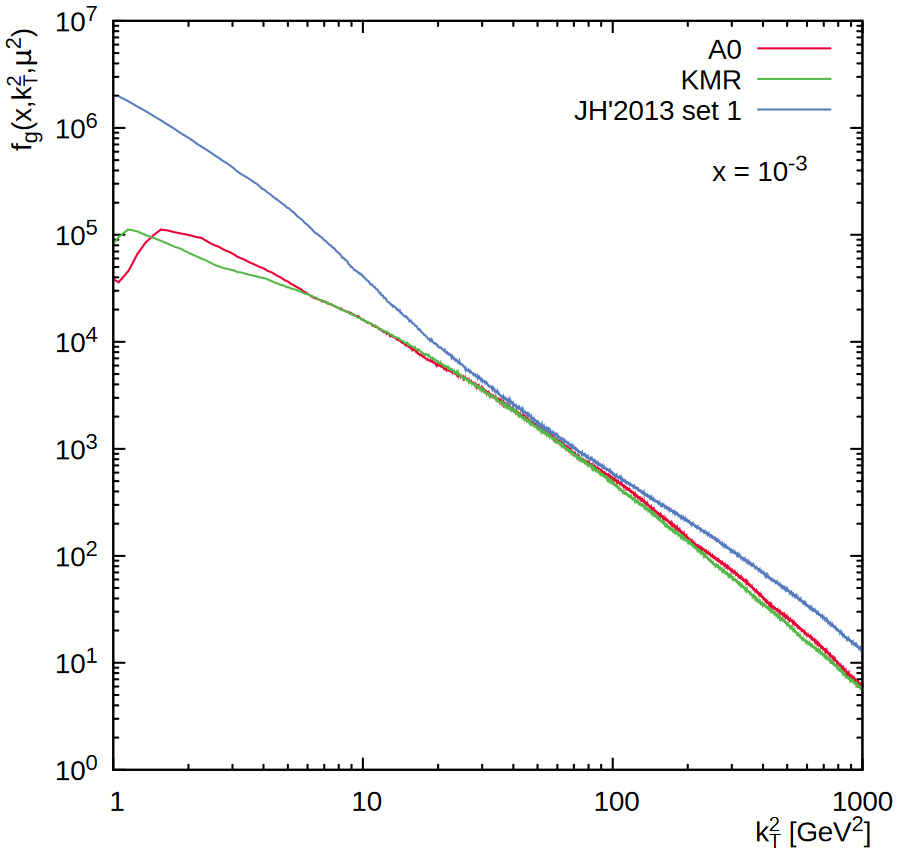}
\caption{The TMD gluon densities in the proton calculated as 
a function of the gluon transverse momentum ${\mathbf k}_T^2$
at different longitudinal momentum fractions $x = 10^{-2}$ (left panel) 
or $x = 10^{-3}$ (right panel) and $\mu^2 = 10^{4}$~GeV$^2$.}
\label{fig1}
\end{center}
\end{figure}

\begin{figure}
\begin{center}
\includegraphics[width=7.0cm]{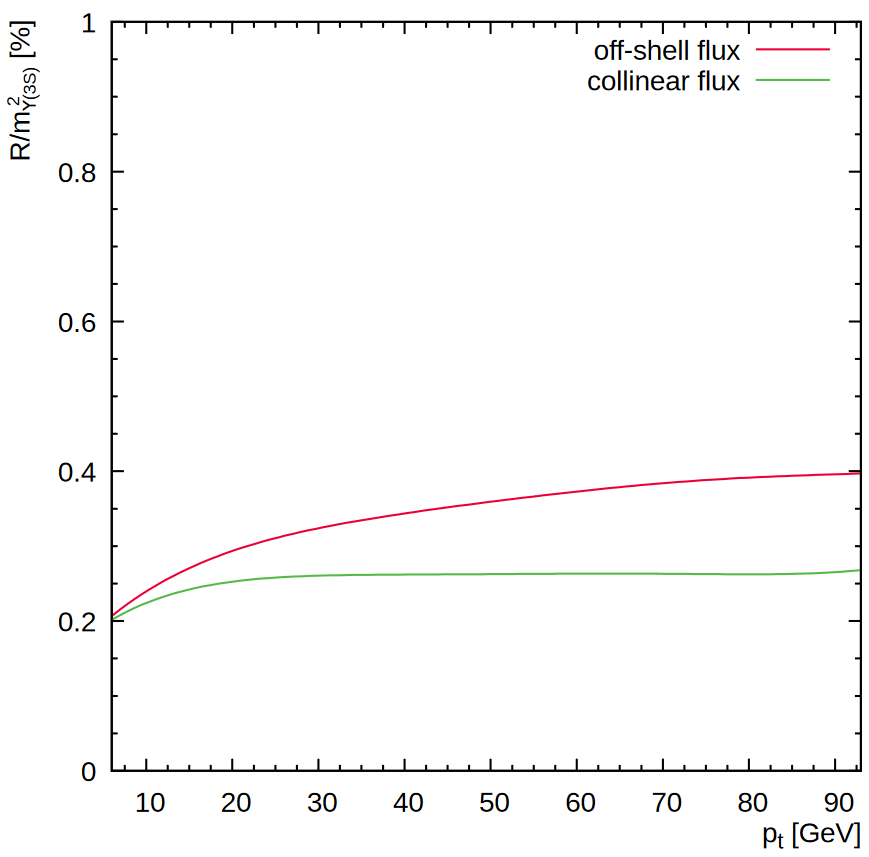}
\includegraphics[width=6.85cm]{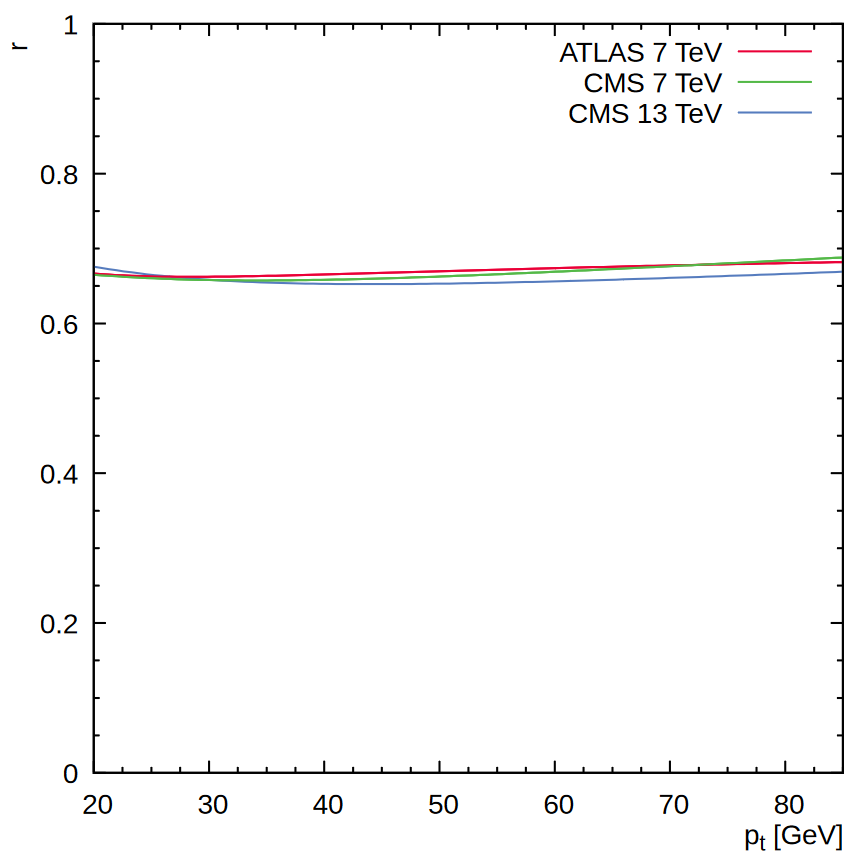}
\caption{The production ratios $R$ (left panel) and $r$ (right panel) 
calculated as a function of $\Upsilon(3S)$ transverse momentum $p_T$ in the 
different kinematical regions.}
\label{fig2}
\end{center}
\end{figure}

\begin{figure}
\begin{center}
\includegraphics[width=7.0cm]{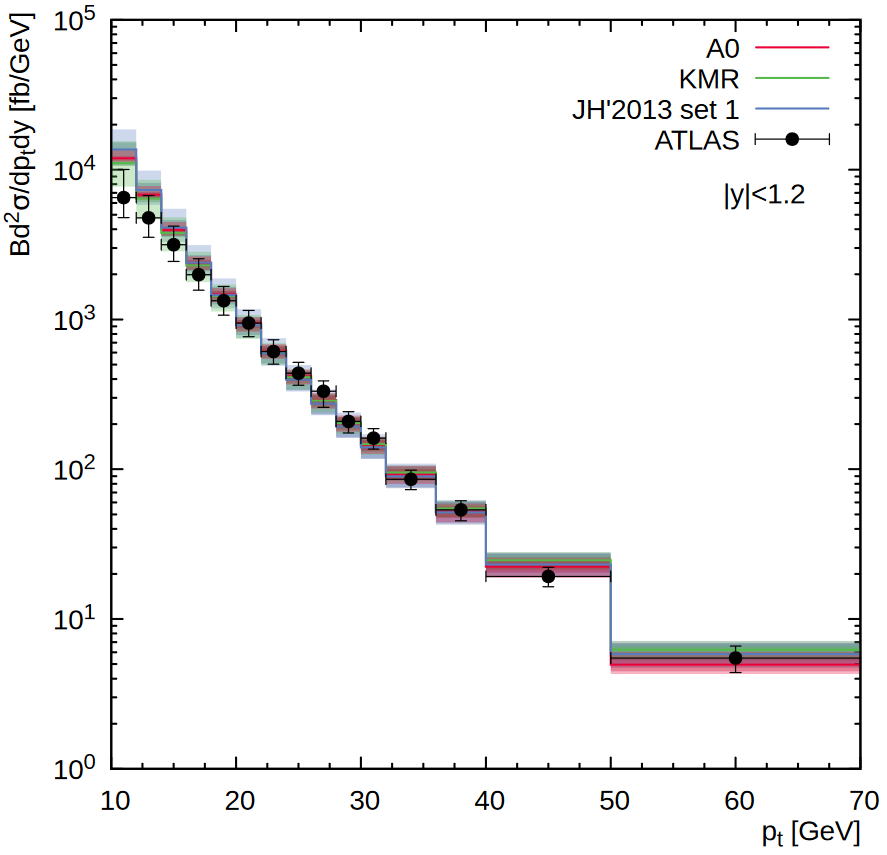}
\includegraphics[width=7.0cm]{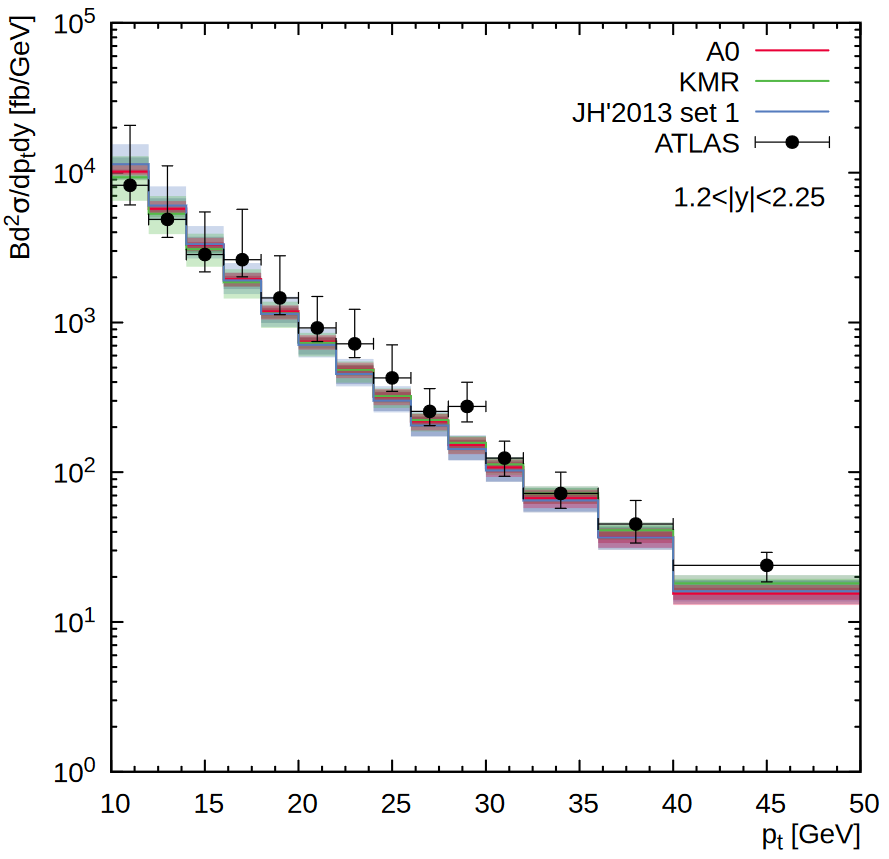}
\caption{Transverse momentum distribution of 
  inclusive $\Upsilon(3S)$ production calculated at 
  $\sqrt s = 7$~TeV in the different rapidity regions. 
  The red, green and blue histograms
  correspond to the predictions obtained with A0, KMR and JH'2013 set 1
  gluon densities. Shaded bands represent the total uncertainties 
  of our calculations, as it is described in text.
  The experimental data are from ATLAS\cite{25}.}
\label{fig3}
\end{center}
\end{figure}

\begin{figure}
\begin{center}
\includegraphics[width=7.0cm]{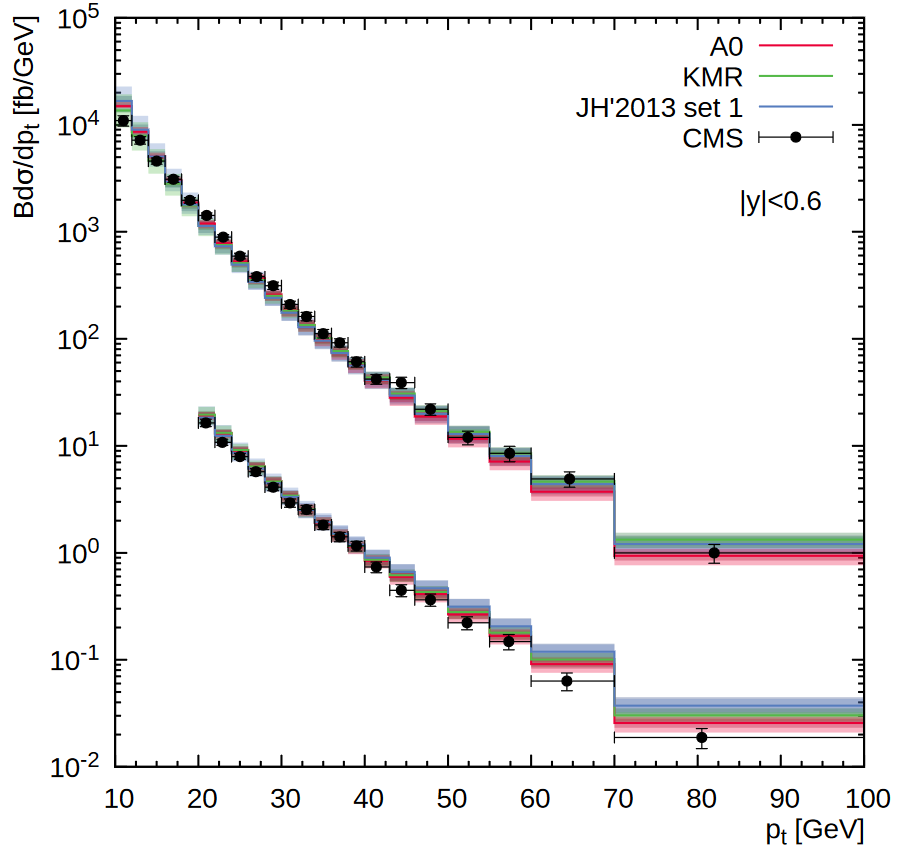}
\includegraphics[width=7.0cm]{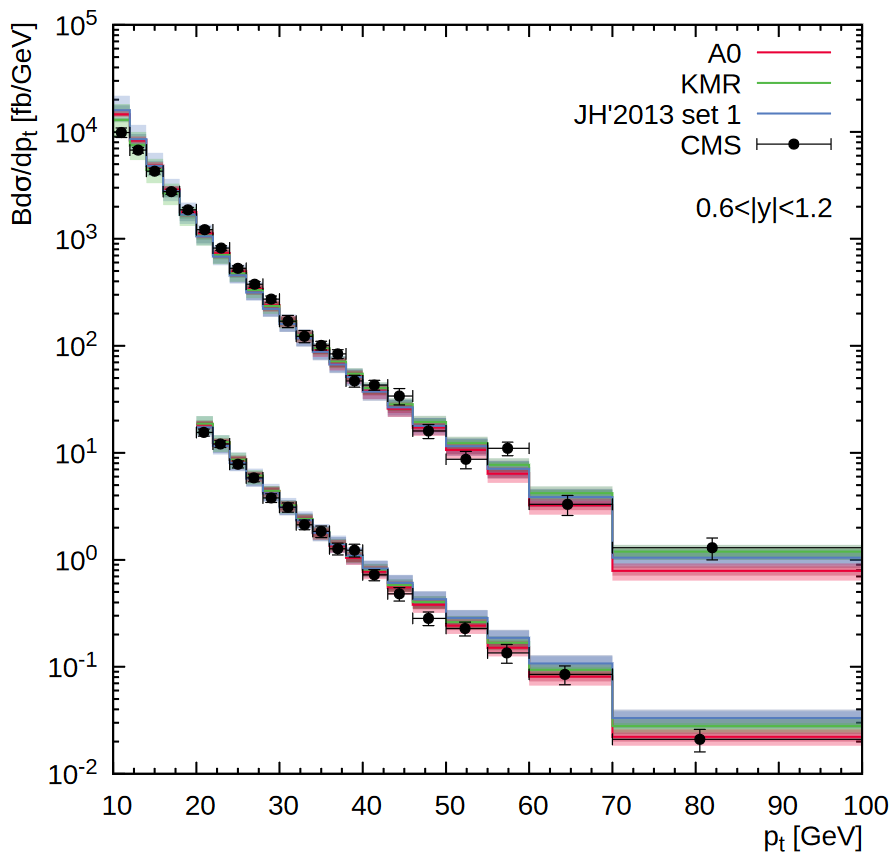}
\includegraphics[width=7.0cm]{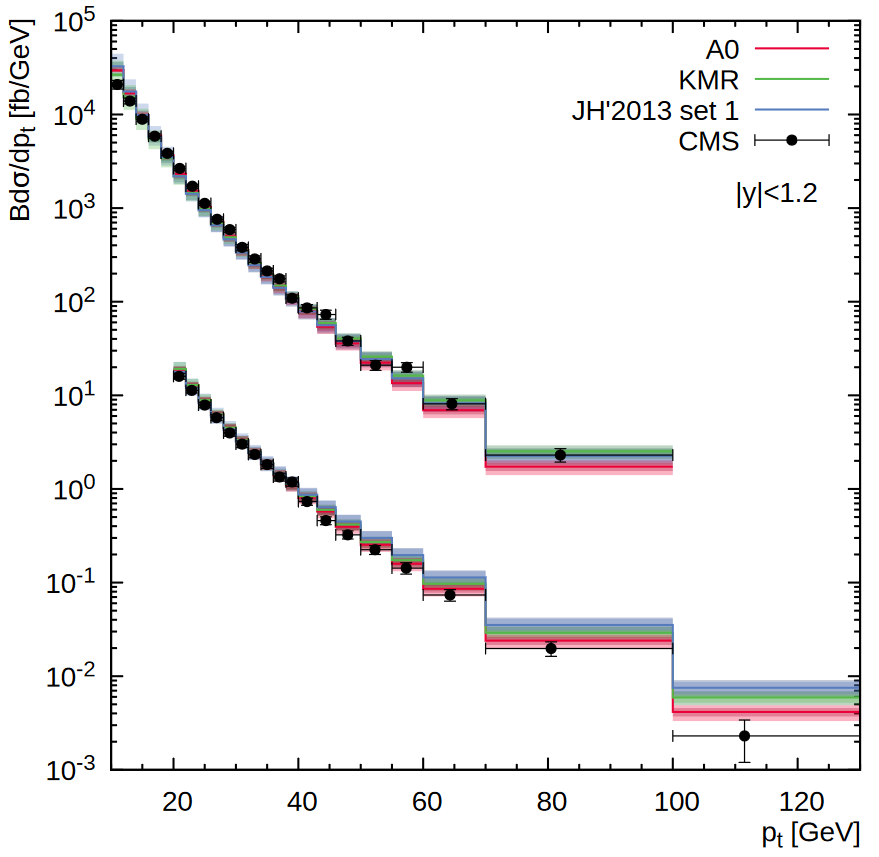}
\caption{Transverse momentum distribution of 
  inclusive $\Upsilon(3S)$ production calculated at $\sqrt s = 7$~TeV
  (upper histograms) and $\sqrt s = 13$~TeV (lower histograms, 
  divided by $100$) in the different rapidity regions. 
  Notation of all histograms is the same as in Fig.~3.
  The experimental data are from CMS\cite{23,24}.}
\label{fig4}
\end{center}
\end{figure}

\begin{figure}
\begin{center}
\includegraphics[width=7.0cm]{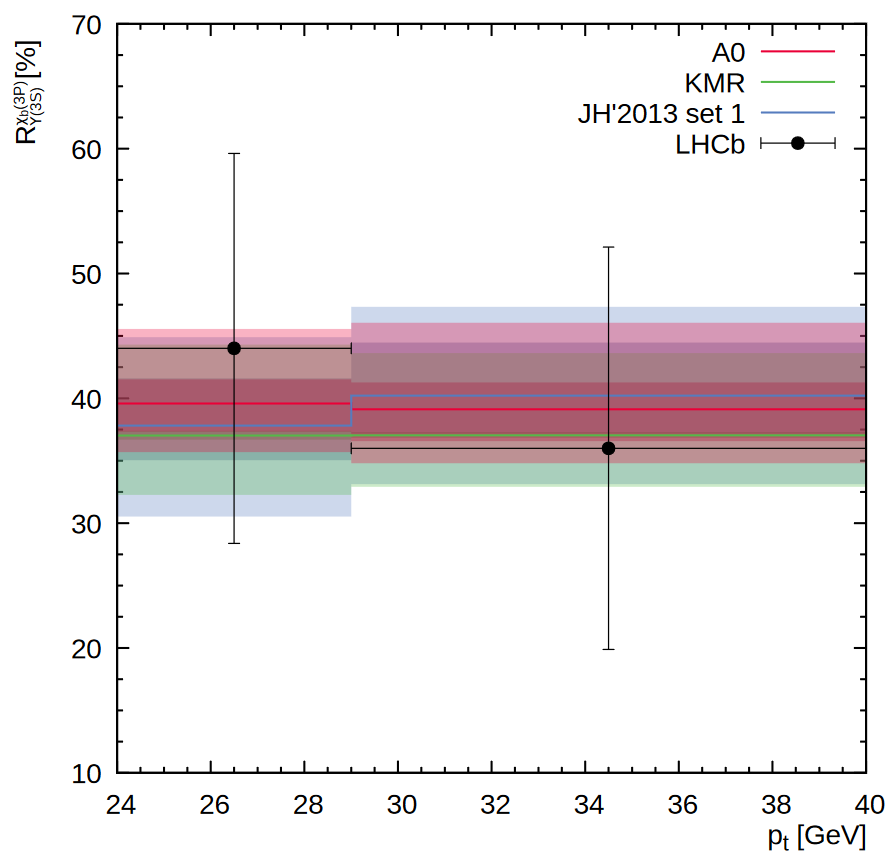}
\includegraphics[width=7.0cm]{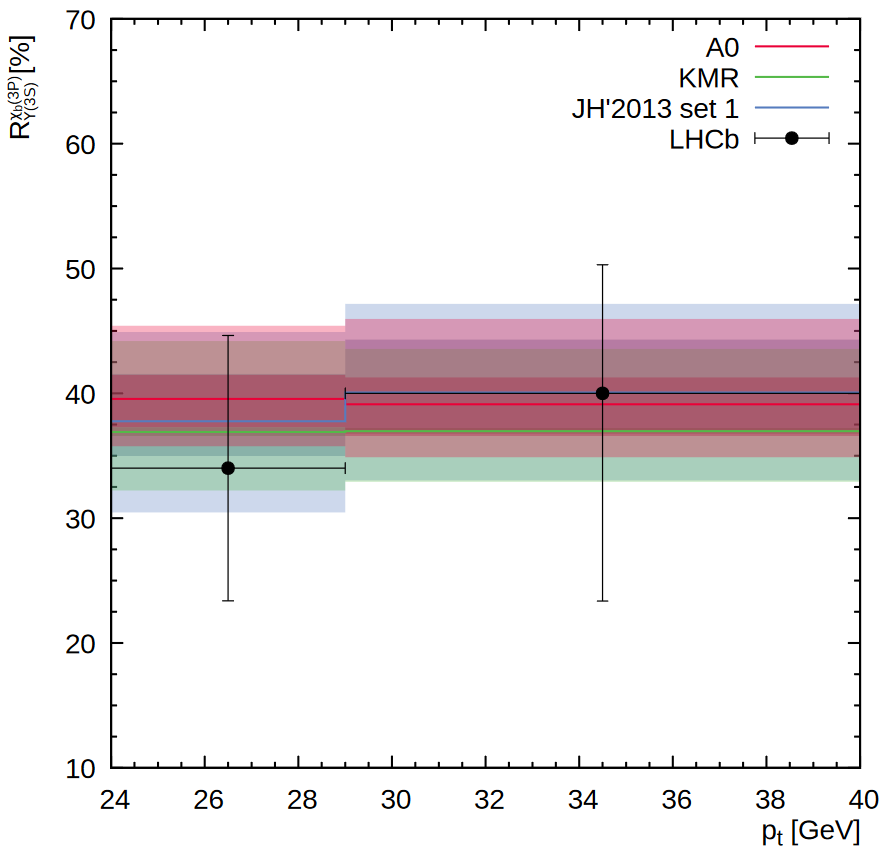}
\caption{The ratio $R^{\chi_b(3P)}_{\Upsilon(3S)}$ 
  calculated as function of $\Upsilon(3S)$ transverse momentum 
  at $\sqrt{s} = 7$~TeV (left panel) and $\sqrt{s} = 8$~TeV (right panel). 
  Notation of all histograms is the same as in Fig.~3.
  The experimental data are from LHCb\cite{35}.}
\label{fig5}
\end{center}
\end{figure}

\begin{figure}
\begin{center}
\includegraphics[width=7.0cm]{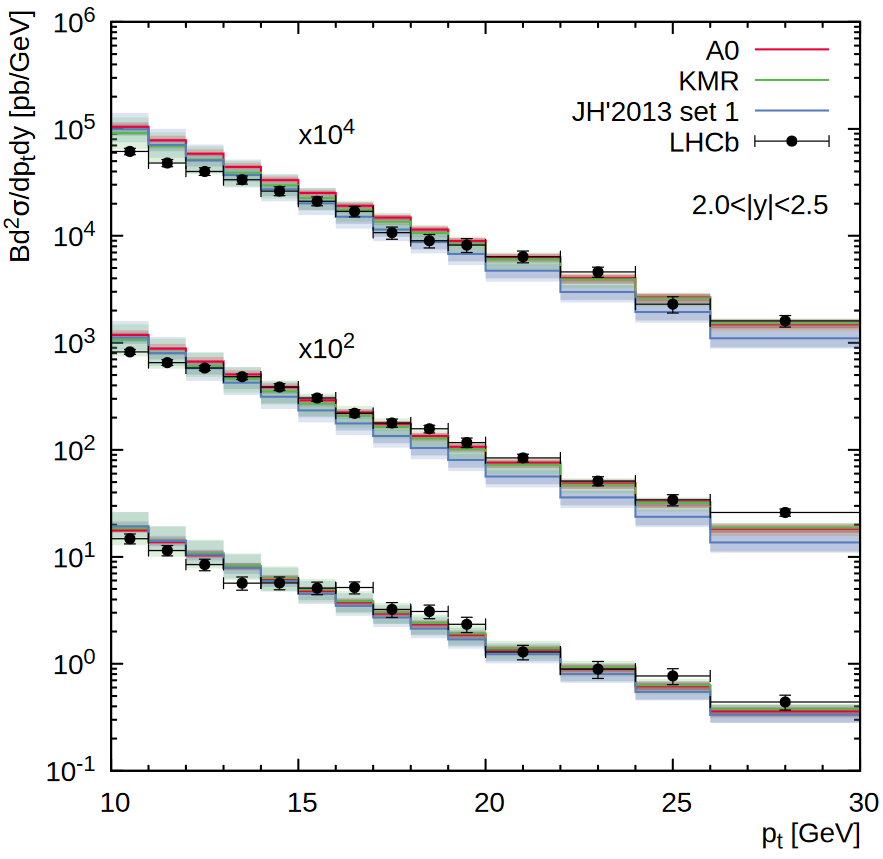}
\includegraphics[width=7.0cm]{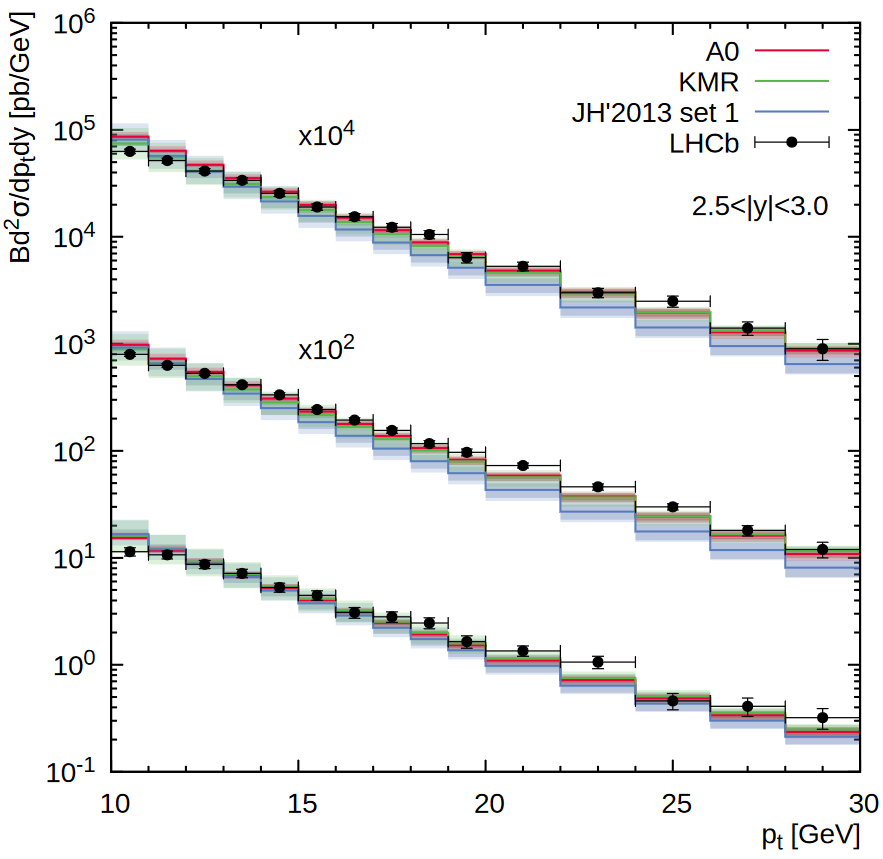}
\includegraphics[width=7.0cm]{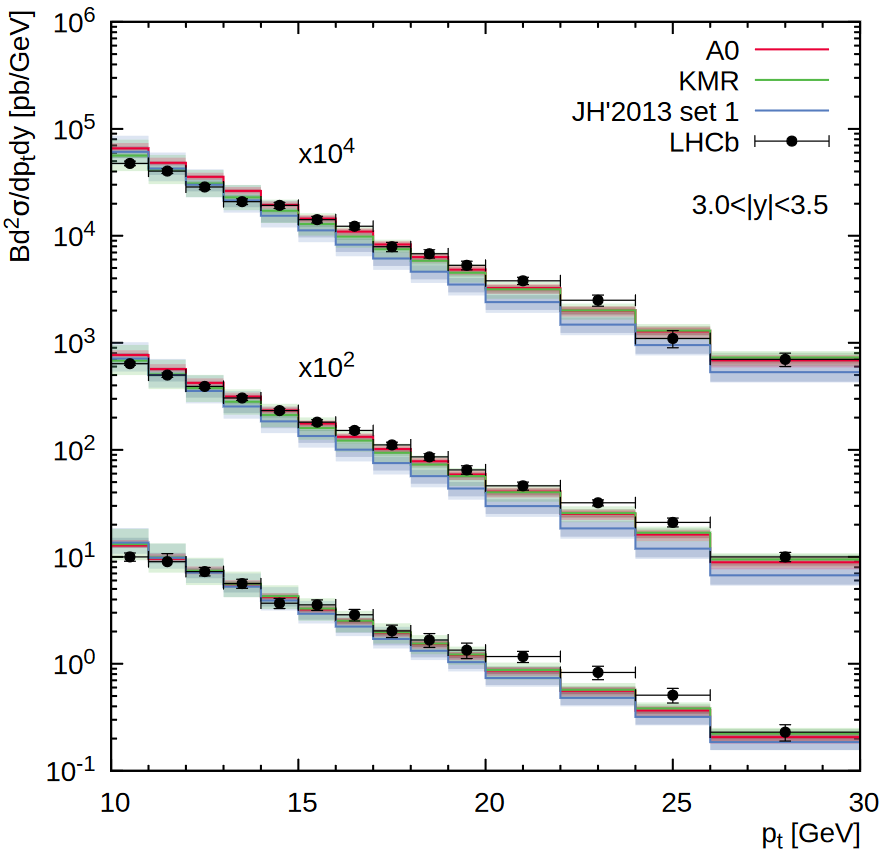}
\includegraphics[width=7.0cm]{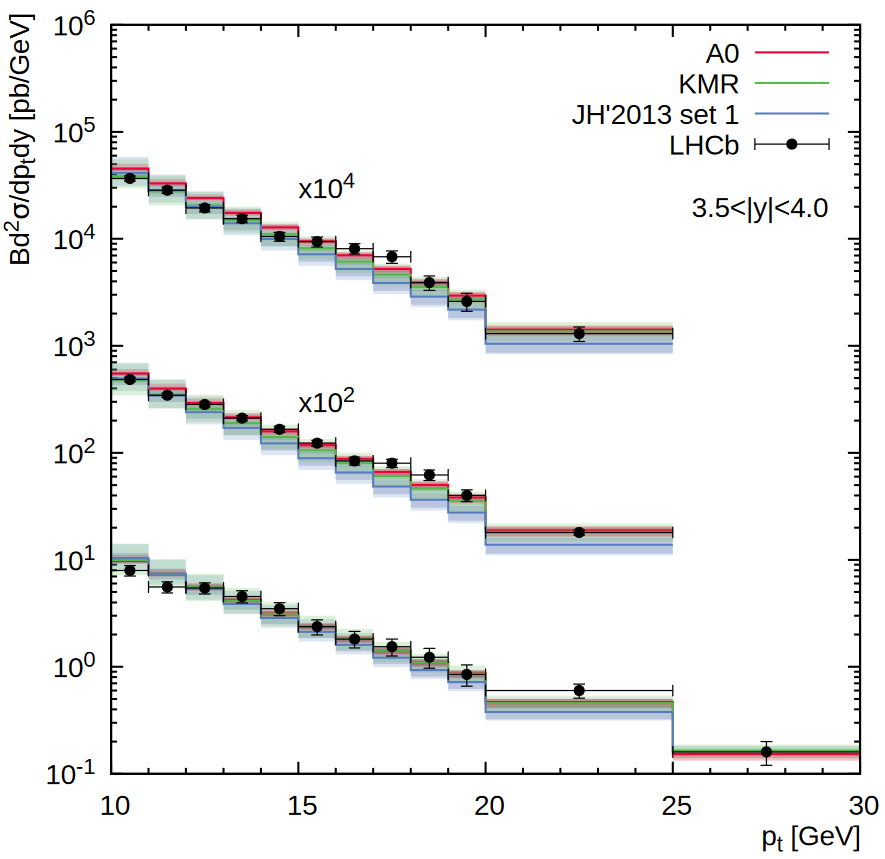}
\includegraphics[width=7.0cm]{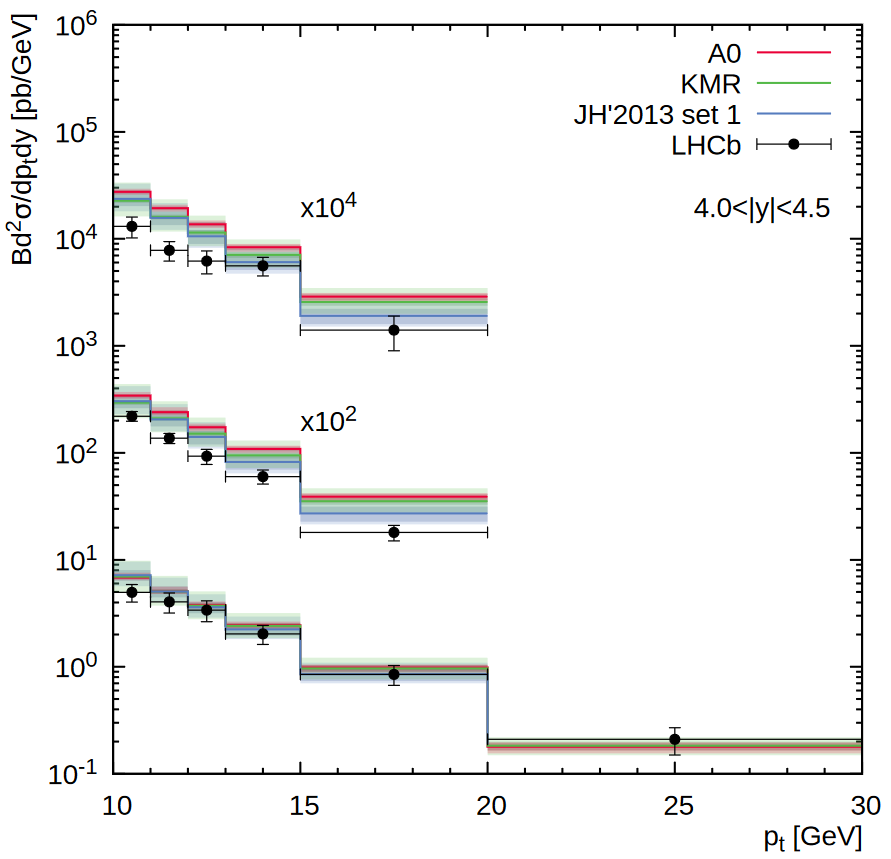}
\includegraphics[width=7.0cm]{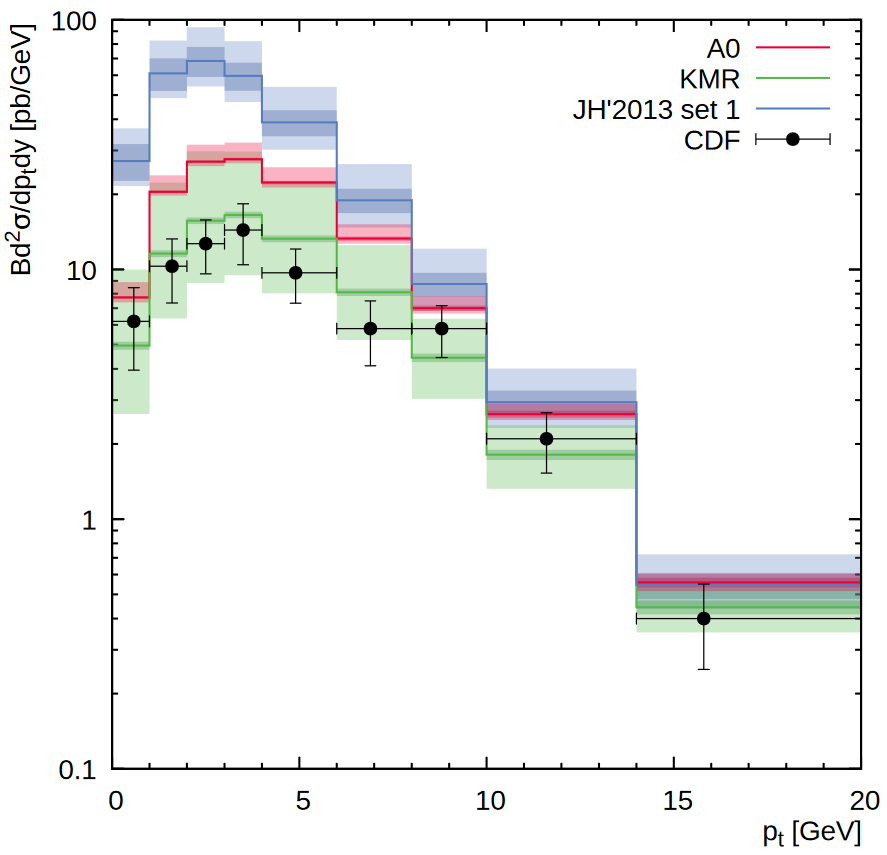}
\caption{Transverse momentum distribution of 
  inclusive $\Upsilon(3S)$ production calculated at $\sqrt s = 1.8$, 
  $7$, $8$ and $13$~TeV in the different rapidity regions. 
  Notation of all histograms is the same as in Fig.~3.
  The experimental data are from CDF\cite{41} and LHCb\cite{26,27}.}
\label{fig6}
\end{center}
\end{figure}

\begin{figure}
\begin{center}
\includegraphics[width=7.0cm]{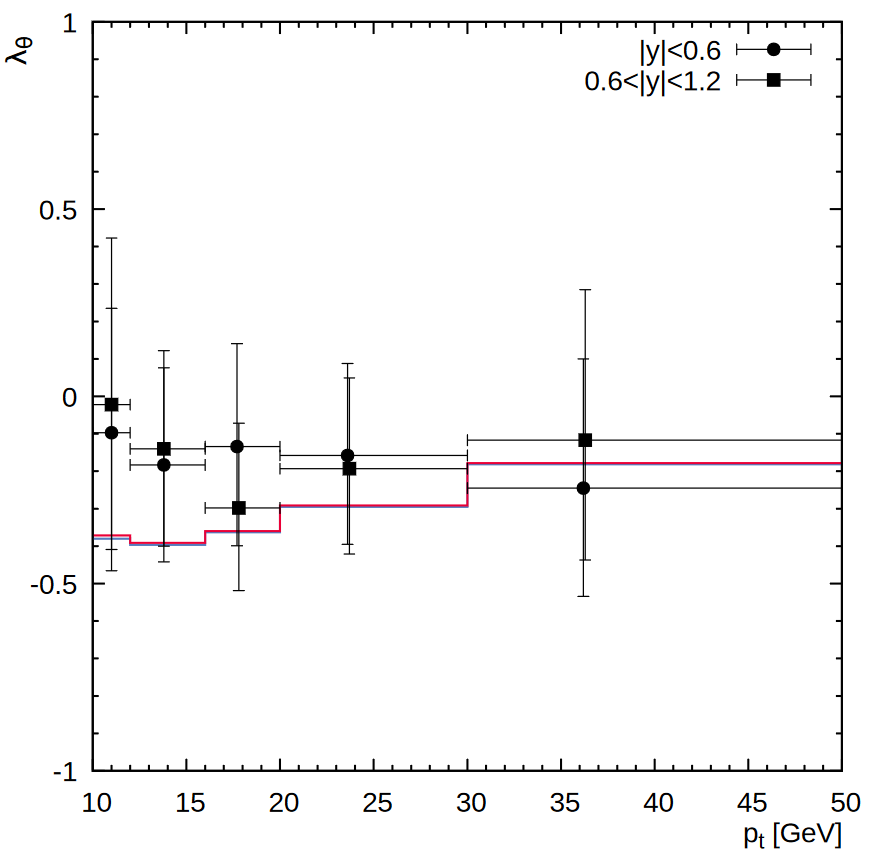}
\includegraphics[width=7.0cm]{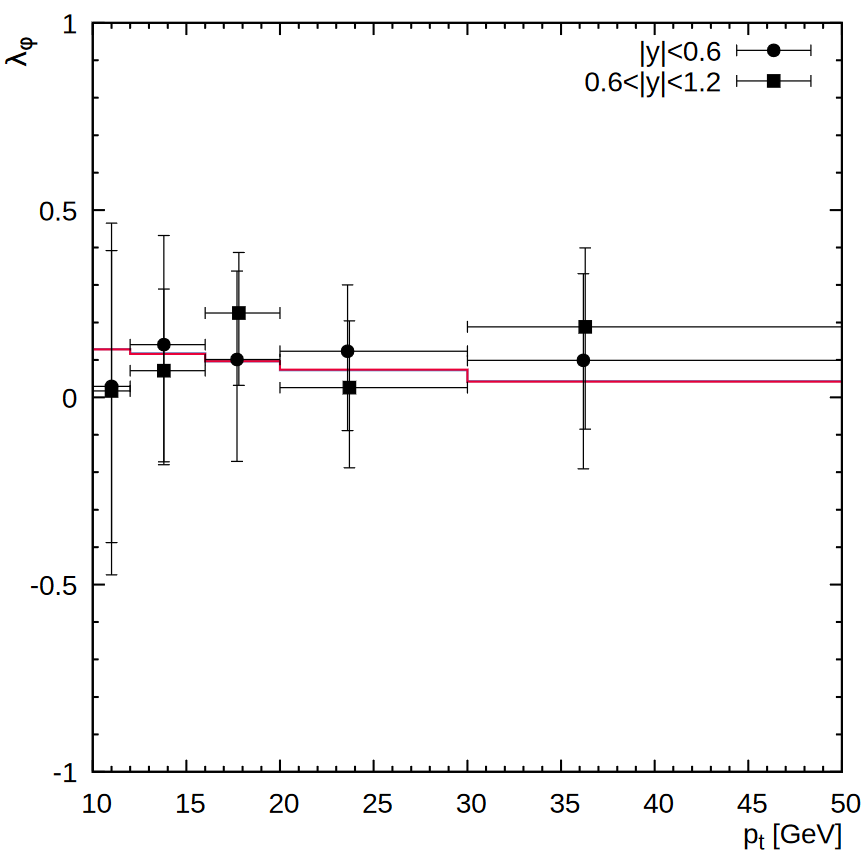}
\includegraphics[width=7.0cm]{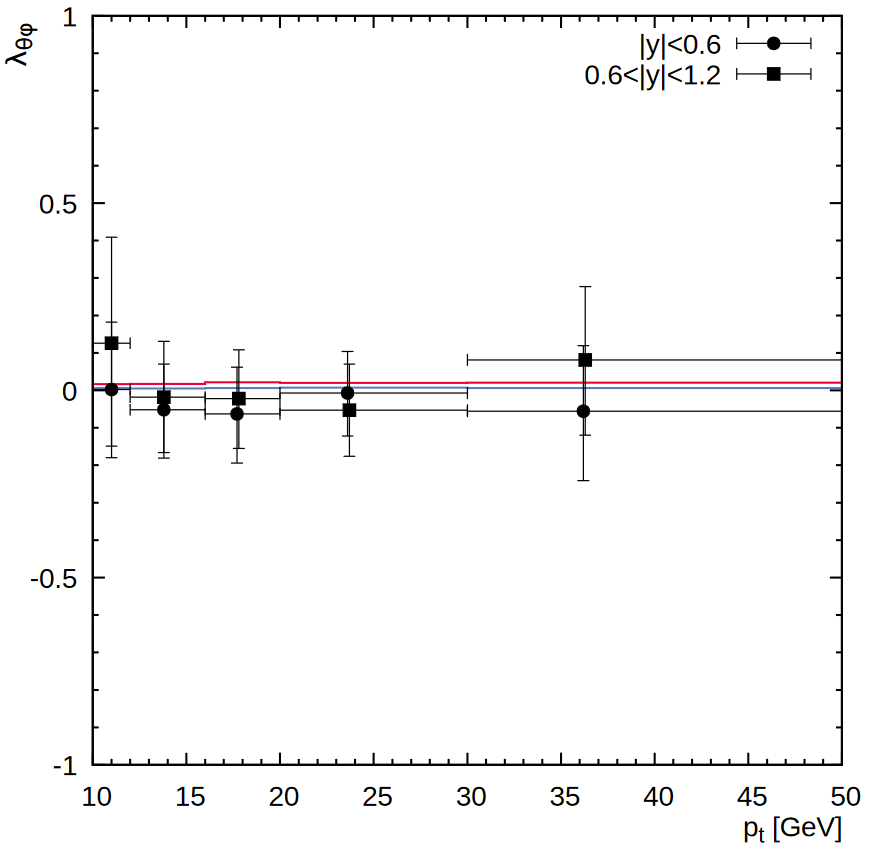}
\includegraphics[width=7.0cm]{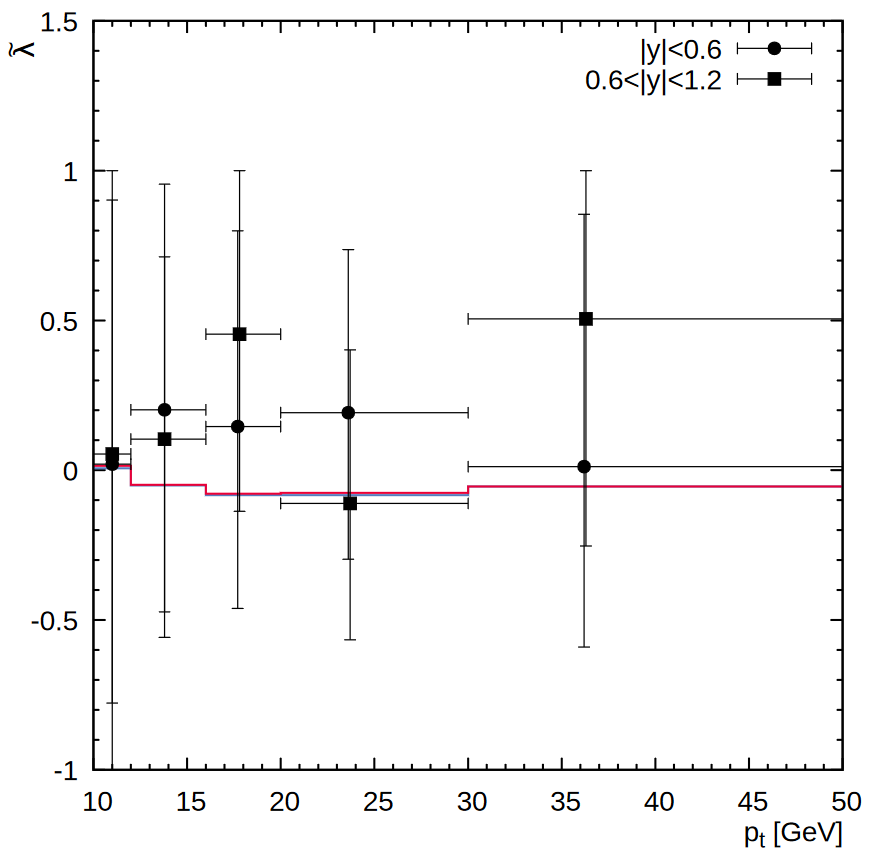}
\caption{The polarization parameters $\lambda_\theta$, 
  $\lambda_\phi$, $\lambda_{\theta\phi}$ and $\tilde\lambda$ of 
  $\Upsilon(3S)$ mesons calculated in the CS frame as function
  of its transverse momentum at $\sqrt{s} = 7$ TeV. 
  The A0 gluon density is used. The blue and red histograms 
  correspond to the predictions obtained at $|y|<0.6$ and 
  $0.6<|y|<1.2$, respectively. 
  The experimental data are from CMS\cite{28}.}
\label{fig7}
\end{center}
\end{figure}

\begin{figure}
\begin{center}
\includegraphics[width=7.0cm]{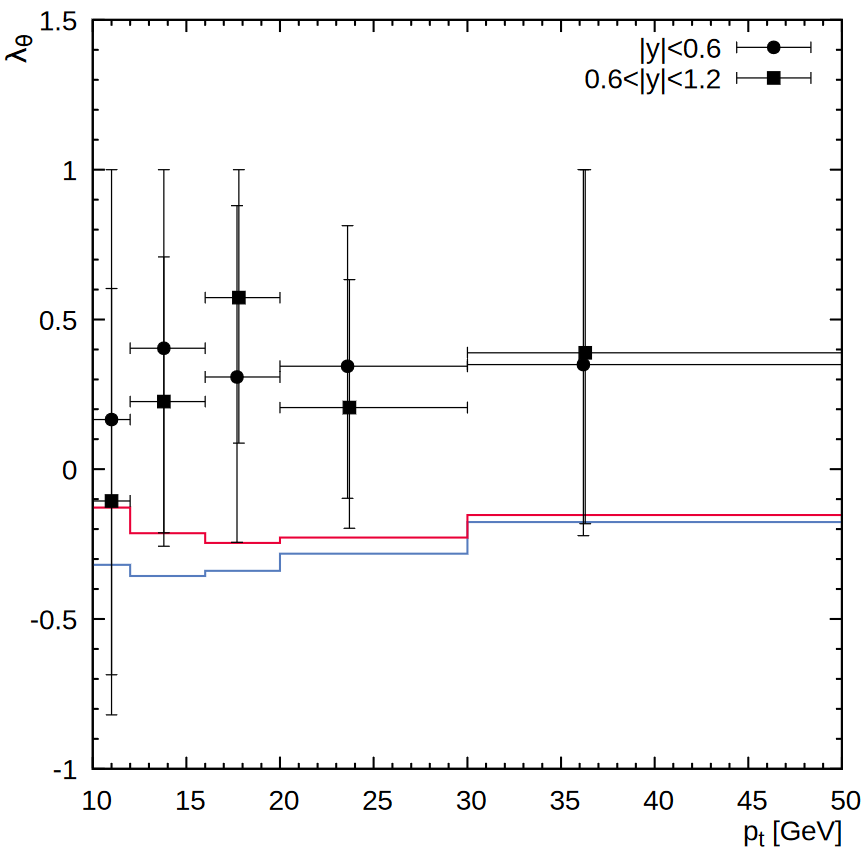}
\includegraphics[width=7.0cm]{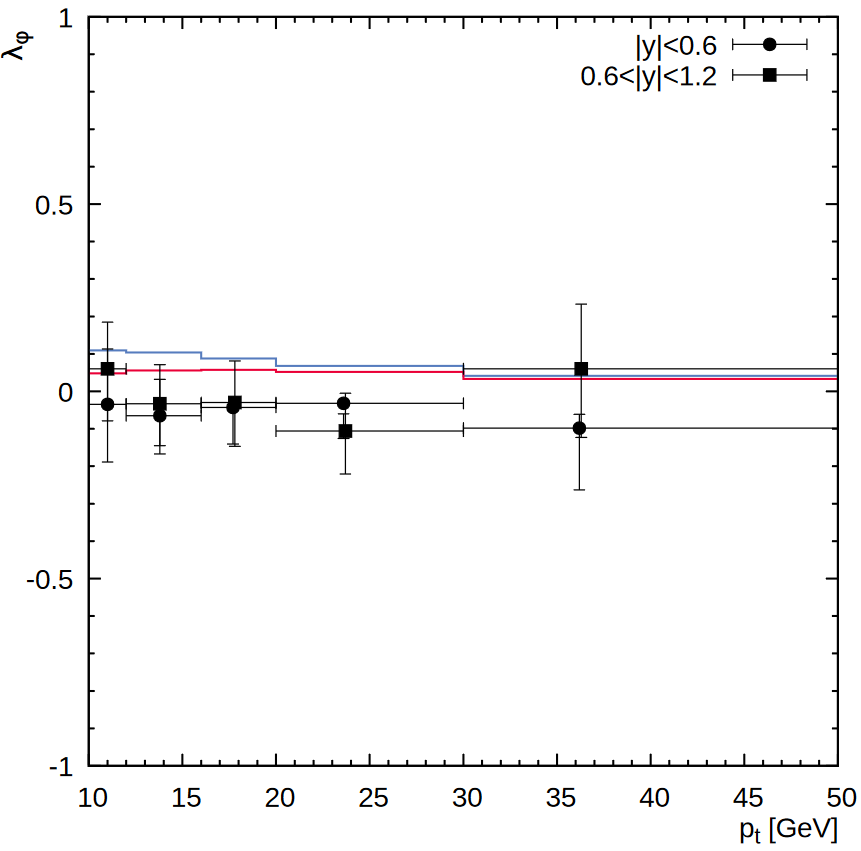}
\includegraphics[width=7.0cm]{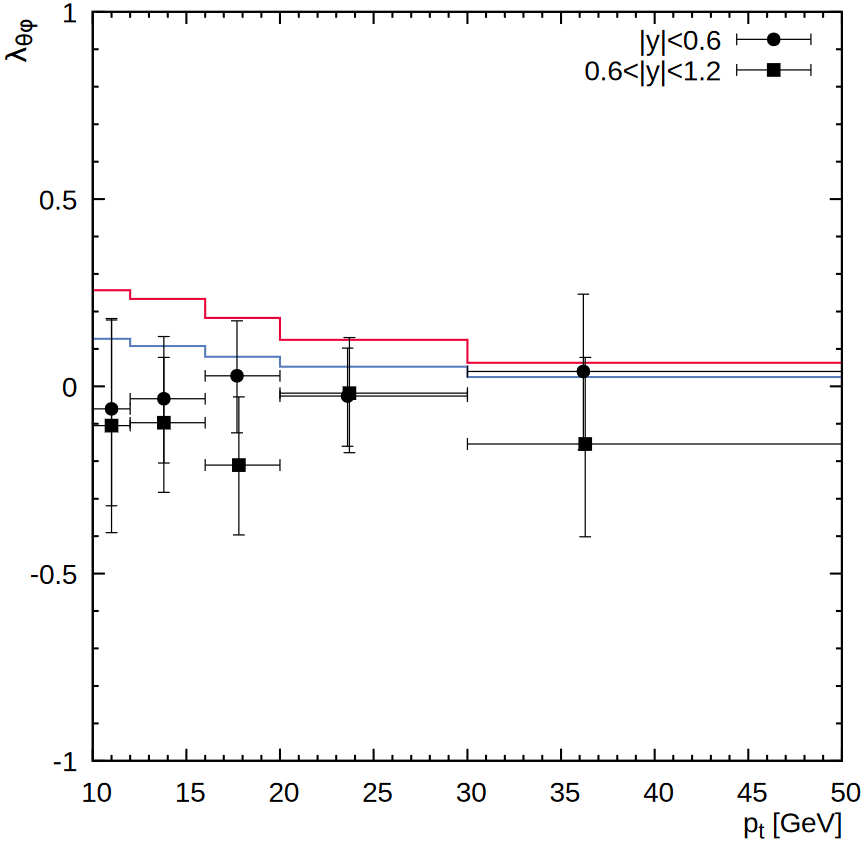}
\includegraphics[width=7.0cm]{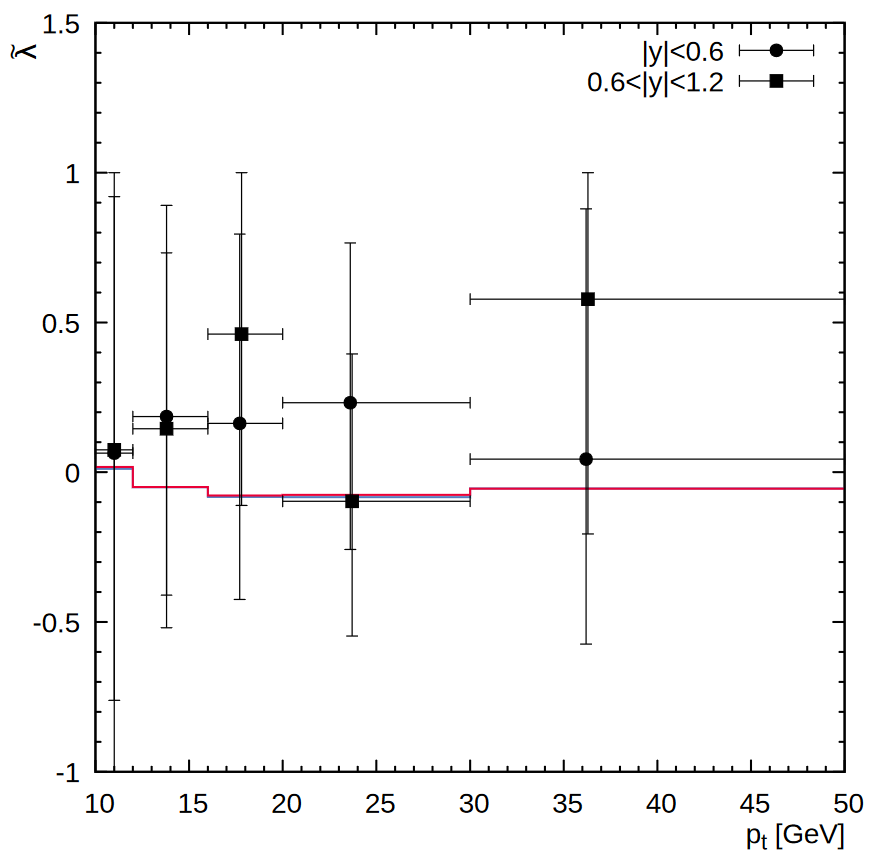}
\caption{The polarization parameters $\lambda_\theta$, 
  $\lambda_\phi$, $\lambda_{\theta\phi}$ and $\tilde\lambda$ of 
  $\Upsilon(3S)$ mesons calculated in the helicity frame as function
  of its transverse momentum at $\sqrt{s} = 7$ TeV. 
  Notation of all histograms is the same as in Fig.~7.
  The experimental data are from CMS\cite{28}.}
\label{fig8}
\end{center}
\end{figure}

\begin{figure}
\begin{center}
\includegraphics[width=7.0cm]{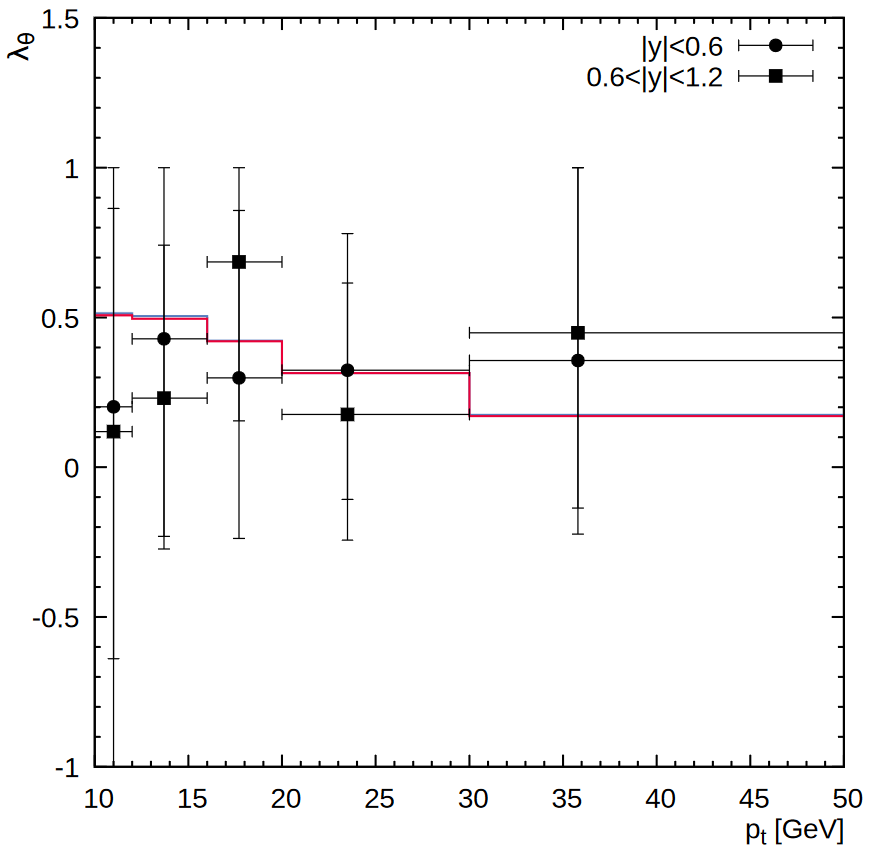}
\includegraphics[width=7.0cm]{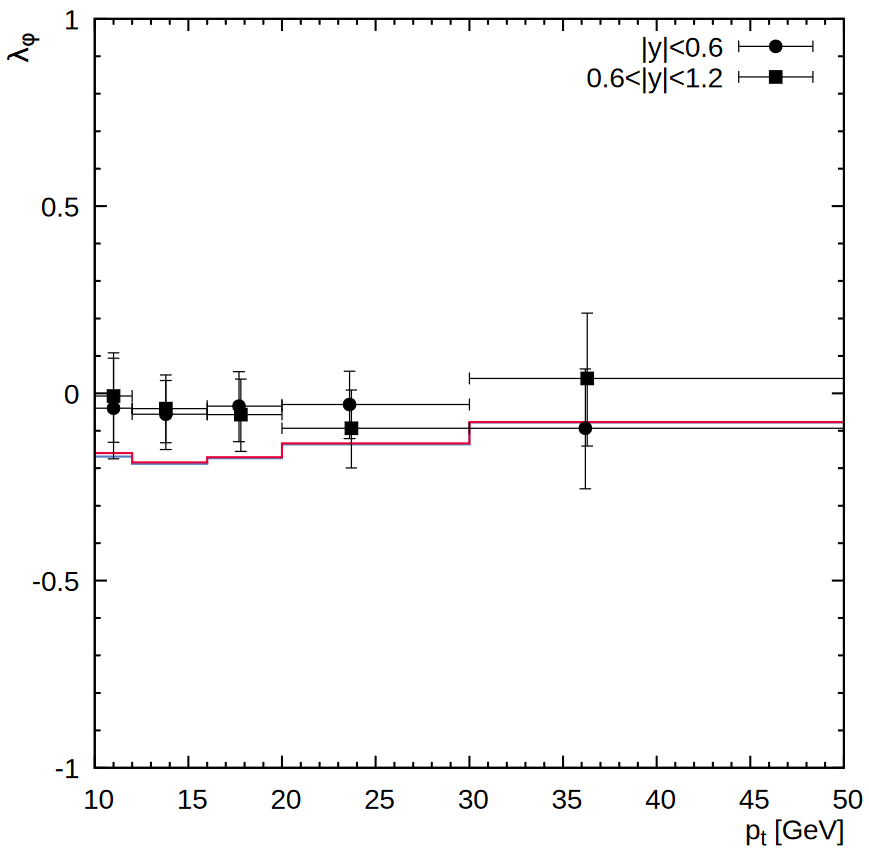}
\includegraphics[width=7.0cm]{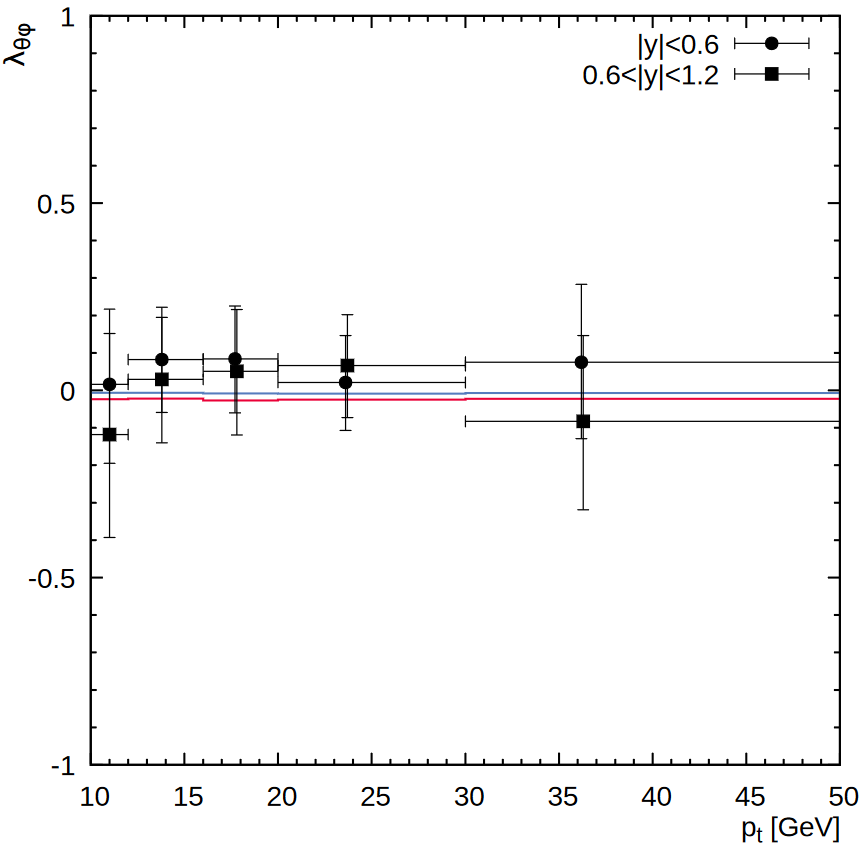}
\includegraphics[width=7.0cm]{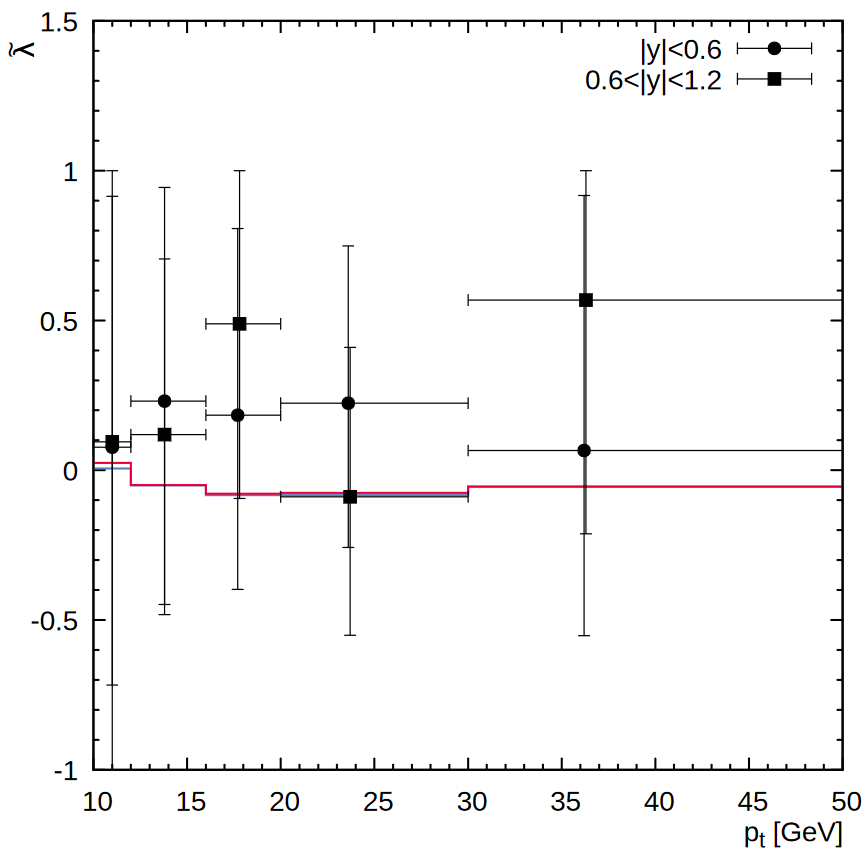}
\caption{The polarization parameters $\lambda_\theta$, 
  $\lambda_\phi$, $\lambda_{\theta\phi}$ and $\tilde\lambda$ of 
  $\Upsilon(3S)$ mesons calculated in the perpendicular 
  helicity frame as function
  of its transverse momentum at $\sqrt{s} = 7$ TeV. 
  Notation of all histograms is the same as in Fig.~7.
  The experimental data are from CMS\cite{28}.}
\label{fig9}
\end{center}
\end{figure}

\begin{figure}
\begin{center}
\includegraphics[width=7.0cm]{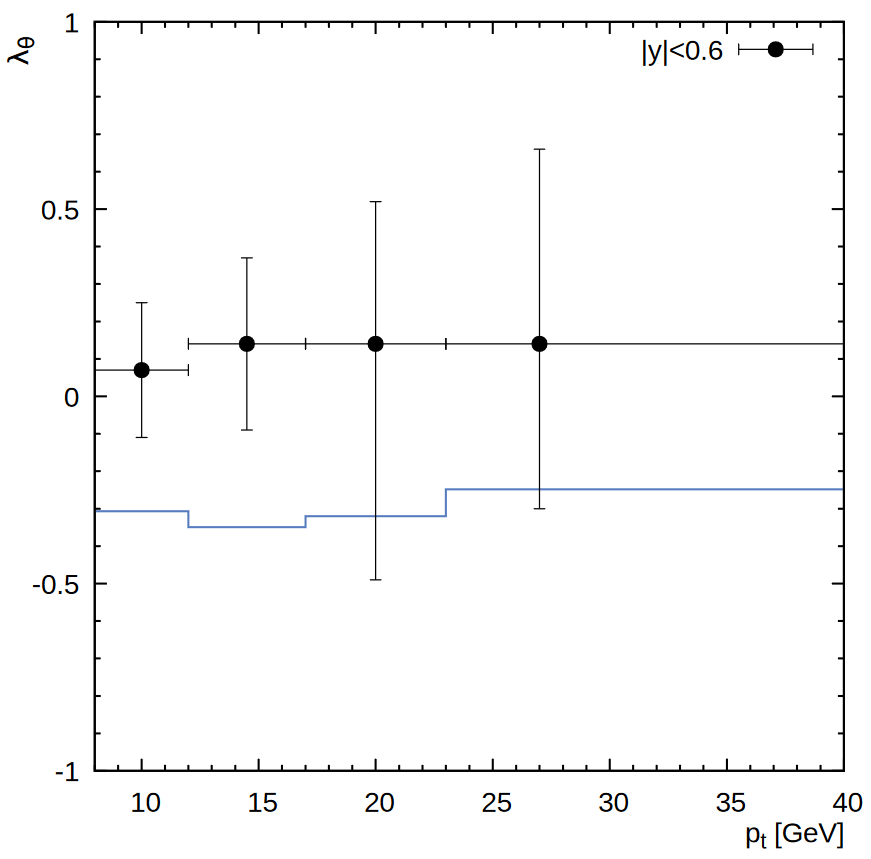}
\includegraphics[width=7.0cm]{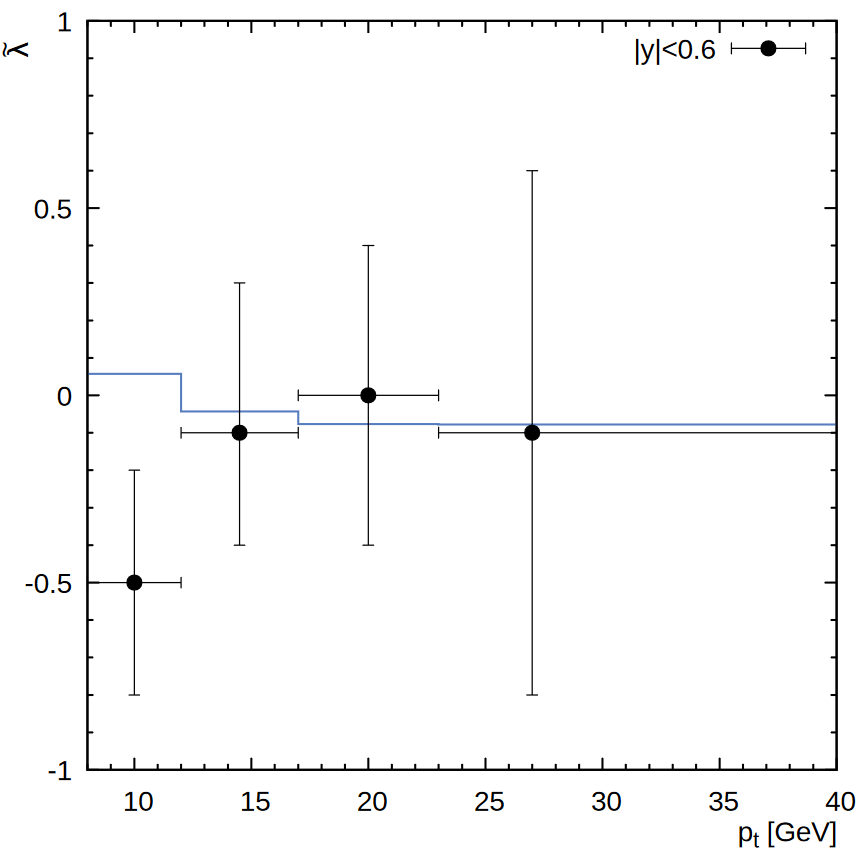}
\caption{The polarization parameters $\lambda_\theta$ and 
$\tilde\lambda$ of $\Upsilon(3S)$ mesons calculated in the 
helicity frame as function of its transverse momentum at 
$\sqrt{s} = 1.96$ TeV. Notation of all histograms is the same as in Fig.~7.
  The experimental data are from CDF\cite{42}.}
\label{fig10}
\end{center}
\end{figure}

\end{document}